\documentclass{article}

 \PassOptionsToPackage{numbers, compress}{natbib}

\usepackage[main, final]{iaseai26}

\usepackage[utf8]{inputenc} 
\usepackage[T1]{fontenc}    
\usepackage{hyperref}       
\usepackage{xurl}            
\usepackage{booktabs}       
\usepackage{amsfonts}       
\usepackage{nicefrac}       
\usepackage{microtype}      
\usepackage{xcolor}         
\usepackage{graphicx}
\usepackage{tabularx}
\usepackage{float}
\usepackage{rotating}

\title{Clear, Compelling Arguments: Rethinking the Foundations of Frontier AI Safety Cases}

\author{%
  Shaun Feakins\\
  UKRI Centre for Doctoral Training in Safe AI Systems (SAINTS) \\
  Institute for Safe Autonomy\\
  Deramore Ln, York YO10 5GH \\
  \texttt{shaun.feakins@york.ac.uk} \\
   \And
  Ibrahim Habli\\
  UKRI Centre for Doctoral Training in Safe AI Systems (SAINTS) \\
  Institute for Safe Autonomy\\
  Deramore Ln, York YO10 5GH \\
   \texttt{ibrahim.habli@york.ac.uk} \\
   \AND
  Phillip Morgan \\
  UKRI Centre for Doctoral Training in Safe AI Systems (SAINTS) \\
  The York Law School \\
  Freboys Ln, York YO10 5GD \\
}

\begin{document}

\maketitle

\begin{abstract}
    This paper contributes to the nascent debate around safety cases for frontier AI systems. Safety cases are structured, defensible arguments that a system is acceptably safe to deploy in a given context. Historically, they have been used in safety-critical industries, such as aerospace, nuclear or automotive. As a result, safety cases for frontier AI have risen in prominence, both in the safety policies of leading frontier developers and in international research agendas proposed by leaders in generative AI, such as the Singapore Consensus on Global AI Safety Research Priorities and the International AI Safety Report. This paper appraises this work. We note that research conducted within the alignment community which draws explicitly on lessons from the assurance community has significant limitations. We therefore aim to rethink existing approaches to alignment safety cases. We offer lessons from existing methodologies within safety assurance and outline the limitations involved in the alignment community’s current approach. Building on this foundation, we present a case study for a safety case focused on Deceptive Alignment and CBRN capabilities, drawing on existing, theoretical safety case “sketches” created by the alignment safety case community. Overall, we contribute holistic insights from the field of safety assurance via rigorous theory and methodologies that have been applied in safety-critical contexts. We do so in order to create a better foundational framework for robust, defensible and useful safety case methodologies which can help to assure the safety of frontier AI systems.
\end{abstract}

\section{Introduction}

Safety cases are used to make clear, defensible arguments that a system is acceptably safe in a given context. This paper considers the nascent approach to safety cases by those involved in and concerned about frontier AI systems, which we term “alignment safety cases”. We argue that alignment safety cases diverge significantly from many foundational methods found within safety-critical systems methodologies, limiting their effectiveness. We present a critical appraisal of the existing approach to alignment safety cases. We aim to recentre the debate using foundational and technology-agnostic methodologies from the field of safety assurance.

Safety cases have traditionally been concerned with assuring safety-critical systems. Safety-critical systems are those systems whose failure could result in loss of life, significant property damage or damage to the environment \cite{knight_2002_safety}. Safety assurance focuses on how to communicate, assess and establish confidence in sufficient risk reduction in high-criticality domains, such as aerospace, automotive, nuclear and chemical contexts \cite{habli_2025_the}. The field, also known as safety science or systems safety, incorporates research areas such as engineering \cite{burton_2019_mind}, human factors within teams \cite{sandom_2002_human}, organisational culture \cite{rae_2020_a}, and safety argumentation \cite{graydon_2017_the}. Safety cases have been used across safety-critical industries for decades. They are best understood within safety science as structured frameworks for thinking about the safety of a system, which should result in a compelling and defensible argument about the system from development through to post-deployment.

Alignment safety cases are a growing field of research which aim to use safety cases to help to assure the safety and alignment of frontier AI systems. Frontier AI systems are defined as highly capable general purpose models which \textit{“match or exceed the capabilities present in today’s most advanced models”} \cite{departmentforscienceinnovationandtechnology_2023_capabilities}. Alignment safety cases often include a particular focus on catastrophic risks \cite{buhl_2024_safety, clymer_2024_safety, goemans_2024_safety, korbak_2025_a, balesni_2024_towards} (Appendix B). We define the alignment safety case literature as the current research direction and industry-adopted approach to safety cases by frontier AI developers. \citeauthor{clymer_2024_safety} and \citeauthor{buhl_2024_safety} are examples of initial research in this area, with later work at the U.K. AI Security Institute building directly on \citeauthor{clymer_2024_safety}'s framework \cite{goemans_2024_safety,korbak_2025_a}.

This paper begins by presenting an overview of safety cases within safety-critical systems. It moves to illustrate core issues with the alignment safety case literature’s understanding of and rationale behind safety cases. We then introduce foundational methodologies within safety assurance, and how these might be applied by the alignment safety case community. Finally, we close with a case study illustrating a basic safety argument about two hazardous events - CBRN capabilities and Deceptive Alignment - presented by frontier AI systems, and how a safety case and associated risk assessment might respond to those hazardous events.

\section{Safety-Critical Systems, Safety Cases and Frontier AI Assurance}

An increasing body of work advocates for applying methodologies derived from safety-critical industries and safety assurance to frontier AI \cite{simpson_2025_voluntary, dalrymple_2024_towards, bengio_2025_international}. On Knight’s definition, that safety-critical systems are those systems wherein failure could result in loss of life or significant property damage \cite{knight_2002_safety}, frontier AI systems fit the basic definition of a safety-critical system in some deployment contexts. 

The safety case is a seminal technique within safety-critical systems. Safety cases are accompanied by a rich body of academic and industrial literature and research. A safety case aims to offer \textit{“a structured argument, supported by a body of evidence that provides a compelling, comprehensible and valid case that a system is safe for a given application in a given operating environment.”} \cite{ministryofdefence_2007_defence}

Within safety-critical systems, there are various characterising features of safety cases, involving argument structure \cite{graydon_2017_the}, confidence assessments \cite{graydon_2017_an} and a through-life process \cite{ministryofdefence_2007_defence,habli_2025_the}. Safety case arguments are often presented in graphical notations, such as Goal Structuring Notation (GSN) \cite{kelly_1999_arguing}. Importantly\textbf{, }the safety case is a through-life document, developed from the start of system development through to decommissioning \cite{ministryofdefence_2007_defence}. Safety cases can therefore occasionally become lengthy documents, involving both system level and component level analyses \cite{habli_2025_the}.

The safety case was first introduced formally in the U.K. for the nuclear industry in 1965 \cite{haddoncavekc_the}. An increasing number of industries which aim to assure the safety of their systems have adopted safety cases, from ML-based safety-critical systems within automotive \cite{palin_2010_assurance}, through to healthcare \cite{thehealthfoundation_2023_using}, high-rise building safety \cite{healthandsafetyexecutive_2024_preparing}, and AI systems themselves \cite{habli_2025_the}.

\subsection{Alignment Safety Cases}

This background to safety cases motivated those involved in assuring the safety and alignment of frontier AI systems \cite{clymer_2024_safety, buhl_2024_safety, hilton_2025_safety}. The fact that safety cases have been used in safety-critical settings explicitly form the justification of \citeauthor{buhl_2024_safety} and \citeauthor{clymer_2024_safety}'s use of safety cases:

\citeauthor{clymer_2024_safety}: \textit{“We first introduce the concept of a “safety case,” which is a method of presenting safety evidence used in six industries in the UK (Sujan et al., 2016). A safety case is a structured rationale that a system is unlikely to cause significant harm if it is deployed to a particular setting.“}

\citeauthor{buhl_2024_safety}: \textit{“A safety case is a structured argument, supported by evidence, that a system is safe enough to deploy in a given way (MoD, 2007). Safety cases are used in many safety-critical industries, such as nuclear power, aviation, and autonomous vehicles (The Health Foundation, 2012; Inge, 2007; Sujan et al., 2016)...However, there is still little clarity on what frontier AI safety cases would look like and how they could be used.”}

Since 2024, safety cases for frontier AI have received significant interest. Several significant reports, such as the Singapore Consensus on Global AI Safety Research Priorities and the International AI Safety Report \cite{bengio_2025_international,bengio_2025_the}, use both \citeauthor{clymer_2024_safety} and \citeauthor{buhl_2024_safety} to define safety cases. This work has also been referred to in both Anthropic and Google DeepMind’s inclusion of safety cases in their safety frameworks \cite{anthropic_2024_responsible,googledeepmind_2025_frontier}. Furthermore, the U.K. AI Security Institute (AISI) made safety cases a specific research agenda in 2024 \cite{irving_2024_safety}. Researchers at AISI have since built specific safety case “sketches” and “templates” which have built directly on \citeauthor{clymer_2024_safety}'s argument subcomponents, such as a “cyber inability argument”  \cite{goemans_2024_safety} and a “sketch of an AI control safety case” \cite{korbak_2025_a}. 

Alignment safety case approaches vary significantly from long-established assurance practices in safety-critical industries, despite borrowing the term, the basic idea behind the methodology and the rationale for safety cases. While it is a novel field, alignment safety case arguments have been presented in order to create justifications post development, upon deployment, as to why systems do not display dangerous capabilities \cite{clymer_2024_safety} , alongside continuous monitoring post-deployment. Post-development justification of why a system is safe is only part of a broader aim of a safety case within the assurance community, and we illustrate the limitations of this approach throughout this paper.

The alignment safety case work rarely or insufficiently considers that the safety case is a tool for through-life consideration of how to mitigate eventual harms, hazards and risks. Although the research is novel, illustrated by \citeauthor{hilton_2025_safety}'s open problems \cite{hilton_2025_safety}, we argue that alignment safety case research is proceeding along a research direction which diverges significantly from the safety-critical techniques it aims to borrow.

\subsection{Rationale}

There are multiple significant differences between the alignment safety case approach and the safety assurance approach. One might argue that this arises from the novelty of advanced AI systems. However, this paper suggests that lessons from the safety assurance literature remain appropriate to mitigating issues with existing alignment safety cases.

Importantly, if the fundamental methods on which alignment safety cases are built are deeply divergent from those of the assurance community, then the rationale underpinning \citeauthor{clymer_2024_safety} and \citeauthor{buhl_2024_safety}’s papers for using safety cases breaks down, as both rely on the fact safety cases have been used in safety-critical industries successfully:

\textit{Fact:} Safety cases have been used in safety-critical settings.
  
  \textit{Premise:} Safety-critical settings provide useful scenarios and 
    techniques which can be translated to frontier AI systems.
    
  \textit{Conclusion}: Safety cases should be used to assure frontier AI systems.
  
  \textit{Condition}: If no relevant safety case techniques from safety-critical 
 settings are translated to frontier AI systems, then the premise 
 underpinning the conclusion becomes irrelevant.

The alignment safety case literature clearly intends to incorporate the condition set out in the above reasoning. Every article surveyed in Appendix B draws on safety-critical methodologies or literature in some form: foundational papers rely on established safety case authors within their arguments \cite{hilton_2025_safety, buhl_2024_safety,clymer_2024_safety}. Similarly, the alignment safety case literature generally aims to use safety case notations and references \cite{balesni_2024_towards,goemans_2024_safety,clymer_2025_an,buhl_2025_an,korbak_2025_a,korbak_2025_how}.

We agree with the literature’s aims to incorporate novel techniques to the assurance of frontier AI systems. Nonetheless, the motivation of this paper is to introduce relevant concepts that we believe are directly translatable to the nascent alignment safety case literature. Without this grounding, the underlying motivation for using safety cases becomes irrelevant and the concept of presenting evidence upon deployment for the safety of a system should be rephrased.

\subsection{Foundational Differences between Alignment and Assurance School}

\subsubsection{The Use of Assurance Literature}

Some work within the alignment safety case community uses safety case literature in ways which might be challenged by those in safety assurance. For example, one paper suggests ‘concretising’ safety case arguments into hard standards \cite{clymer_2024_safety}. Hard standards are rigid and static. Safety cases are dynamic documents \cite{denney_2017_dynamic}. Safety cases are not templates which are thereafter hardened into rules. This proposal directly limits the underlying justification of safety cases, which is based in a goal-setting regulatory framework which prioritises flexibility and context-specific safety arguments rather than prescriptive standards \cite{haddoncavekc_the}.

There also appear to be issues within the construction of safety argument notations, a fundamental element of safety case construction \cite{graydon_2017_the}. For example, in one Claims, Arguments, Evidence framework, a claim (C2.2: This AI System poses no risk of novel Cyberattack) is supported by evidence instead of arguments \cite{goemans_2024_safety}. This misses a core aspect of safety case notation, which involves evidence-based explanation and justifications of claims via argumentation \cite{kelly_1999_arguing}. Other approaches base safety case methodologies on ideas which would surprise safety engineers, such as the idea of a ‘national security safety case’ \cite{ortega_2025_ai}. Historically, safety cases and security cases have been separate domains (despite efforts to integrate them), particularly within the defence context of national security. This is encapsulated by \citeauthor{alexander_2017_from}, which examines the application of assurance cases to the security domain, noting in separate sections that “there are practical challenges in moving to security cases” and that “there are significant differences between safety and [traditional] security” \cite{alexander_2017_from}. 

One influential paper briefly suggests reviewing risk cases alongside safety cases \cite{clymer_2024_safety}, a recommendation which has since been directly cited by the International AI Safety Report \cite{bengio_2025_international}. However, risk cases are a method which have been reviewed and rejected by industry authorities, such as the MOD, within safety assurance \cite{thehealthfoundation_2023_using,rae_2017_probative}, and were covered in limited detail within the paper \cite{clymer_2024_safety}. Furthermore, risk cases were initially designed to replace, not complement, safety cases \cite{haddoncavekc_the,rae_2017_probative}. 

These issues demonstrate the importance of stronger collaboration between the safety assurance community and those with expertise and deep familiarity with frontier AI safety.

\subsubsection{Deployment vs development settings}

Many alignment safety cases either refer only to deployment settings or cover development settings in limited detail \cite{clymer_2025_an, goemans_2024_safety, clymer_2024_safety, hilton_2025_safety}. This work has to date omitted substantial discussion of foundational through-life elements within systems safety. This might include altering development conditions in response to certain concerns, such as altering pre-training techniques or post-training methods to limit CBRN capabilities \cite{chen_2025_enhancing}. It might involve management considering more abstract potential harms of developing novel architectures prior to development, such as the risks involved in more performant \cite{hao_2024_training} but obfuscated chain-of-thought reasoning \cite{korbak_2025_chain}. It might also involve the pre-deployment testing outlined by \citeauthor{clymer_2024_safety} and others in the literature, as well as steps taken post-deployment through to decommissioning. While alluded to in the literature \cite{goemans_2024_safety, balesni_2024_towards, hilton_2025_safety}, the focus on through-life evidence has been more limited.

The alignment safety case literature could be viewed as basing its interpretation of safety cases on the idea that a system ‘is safe to deploy’. However, the justification for the system being safe to deploy in this literature is not currently due to through-life assurance, as underpins the safety assurance literature (e.g. ‘safe’ design and ‘safe’ deployment) \cite{ministryofdefence_2007_defence, thehealthfoundation_2023_using,habli_2025_the}. Instead, it is presented as safe because the developer cannot find issues with the system upon deployment. \citeauthor{clymer_2024_safety} discuss the safety case as applying to a specific ‘deployment window’. They silo this approach by noting that they ‘specifically focus on deployment decisions’. While noting that their framework \textit{``could also be adapted to decisions about whether to continue AI training"} \cite{clymer_2024_safety}, this necessarily separates interrelated areas of a model’s lifecycle \cite{clymer_2024_safety}. We believe this justification is insufficient: it would be insufficient in any other safety-critical industry for a model developer to test a model on deployment, without consideration of the broader system’s earlier decisions and processes, and then claim that it is safe to deploy. 

The use of post-development justifications for deployment within the alignment safety case literature indicates a significant divergence from the safety assurance approach. By focussing on deployment decisions, the alignment safety case literature necessarily argues why a system is safe to be released \textit{because it does not do bad things upon deployment}, rather than the fact that a \textit{system is safe because developers have made careful decisions before and throughout the development and deployment lifecycle}. An equivalent distinction may be made between aerospace engines and frontier AI. Within aerospace, systems are safe to deploy because certain engineering decisions have been made throughout the development lifecycle to reduce the chance of physical harm occurring, not because when planes are tested they do not crash. While testing is a part of pre-release, for example by testing the impact of birdstrikes on propellers \cite{downer_2024_rational}, it forms one of many steps within the safety argument lifecycle.

More recent work, such as \citeauthor{hilton_2025_safety}, includes discussion of this issue, but offers minimal detail. \citeauthor{buhl_2024_safety} briefly discuss through-life steps, drawing on best practice from the safety assurance community. \citeauthor{hilton_2025_safety} do mention actionable development steps and engage with the risk assessment process underpinning a safety case, mentioning development and pre-deployment steps within this process. However, this content is left relatively unexplored. For example, \citeauthor{hilton_2025_safety} imply that organisational and training culture arguments are only relevant for ‘low-capability systems’ \cite{hilton_2025_safety}. In contrast, one might argue that the higher the capability of a system, the more likely it is that the system needs to be deployed within safe organisational contexts in order to mitigate catastrophic risks from opaque systems.

Alternatively, \citeauthor{hilton_2025_safety} ask at the end of their paper \textit{“To what extent do current training pipelines incentivize deceptive behaviour?”}. This type of question is pertinent to a safety case, which often deals with uncertainty \cite{nuclearenergyagency_2004_management}. However, \citeauthor{hilton_2025_safety} split safety cases into separate components, such as a ‘pretraining safety case’ \cite{hilton_2025_safety}, which would necessarily split the holistic pipeline of safety evaluation into separate components, rather than acknowledging the complexity of frontier systems and the interrelatedness of development steps. This undermines the strength of a safety case in allowing for a holistic evaluation of a system, rather than individual components of a technology.

The omission in alignment safety case literature of through-life considerations within safety assurance may have had an impact on how safety cases are viewed by those in the wider alignment community. For example, \citeauthor{bowen_2025_ai}, who reference alignment safety case literature and the U.K. Ministry of Defence, effectively strawman the safety case by stating that under the 1997 U.K. Ministry of Defence Standards, \textit{“safety evaluations only need to confirm a single compelling safety case”} \cite{bowen_2025_ai}. This statement could be a response to either alignment safety cases or the idea of a safety case within safety assurance. \citeauthor{bowen_2025_ai} rely on the statement to justify that safety cases are insufficient, as developers only need to present an argument about the model at pre-mitigation or post-mitigation stage, and thereafter \textit{``testing can stop"} \cite{bowen_2025_ai}. This appears to be a questionable interpretation of Ministry of Defence safety assurance requirements. For example, the Ministry of Defence has various comprehensive documents on through-life capability management \cite{defencesafetyauthority_2024_dsa}. Similarly, the safety evaluation of any safety-critical system does not stop upon release. This is among the most basic principles of safety engineering, supported by a rich body of debate and evidence (e.g. criticised by \citeauthor{kelly_2008_are} in 2008 as ‘\textit{Safety case shelf-ware}’ \cite{kelly_2008_are}). When military software or hardware is released, post-deployment strategies for potential issues are constantly reevaluated until decommissioning \cite{defencesafetyauthority_2024_dsa}. It is feasible that \citeauthor{bowen_2025_ai} have interpreted safety cases as they do due to the alignment safety cases conducted within the broader field of alignment research. In either case, this issue illustrates the downstream impacts of a misunderstanding of the foundational aims of safety cases.

We aim to respond to and build on these issues and interpretations of safety assurance perspectives to recentre the debate around safety cases. We offer perspectives from safety assurance to illustrate a deeper outline of a risk assessment underpinning a frontier AI safety case, where there may be overlaps between safety assurance risk assessments and frontier AI risk assessments, and how that might inform a safety argument.

\section{Risk assessment}

Risk management informs and underpins a safety case within safety science. The safety case therefore reflects associated systematic risk management steps. This principle moves safety cases from “Paper Safety” rubber-stamp documents through to a document which helps to assure that systems are safe in a given context. Paper safety represents the idea that when safety cases are done incorrectly, they can simply be confirmatory, bureaucratic exercises which rubber-stamp safety arguments \cite{haddoncavekc_the}. Existing criticism of safety cases within the alignment community indicates well that nascent work on alignment safety cases may not sufficiently illustrate the risk management process throughout the development lifecycle \cite{greenblatt_2025_focus}. This approach could inadvertently fall into the trap of ‘paper safety'.  

This argument is implicit in \citeauthor{greenblatt_2025_focus}’s criticism of safety cases for frontier AI, given he suggests that the approach has bad ‘epistemic effects’ \cite{greenblatt_2025_focus}. \citeauthor{greenblatt_2025_focus} states that safety cases are not useful tools because developers should instead focus on collecting evidence \cite{greenblatt_2025_focus}. In contrast, the focus of a safety case within the safety assurance community should be to collate risk-based evidence throughout the development process. \citeauthor{greenblatt_2025_focus}, whether intentionally or inadvertently given his work within the alignment community, echoes our criticism of alignment safety cases by drawing on well-established debates within safety assurance. Furthermore, \citeauthor{bowen_2025_ai} implicitly make this criticism of alignment safety cases by interpreting a safety case as a justification that ‘confirms’ a single document.

The alignment safety case literature evidently does not aim to produce confirmatory documents, given there is significant rigour in the evaluative methods \cite{clymer_2024_safety,goemans_2024_safety,balesni_2024_towards}. Instead, we suggest that the idea behind producing a safety case for frontier AI systems, as well as the testing methodologies presented by the wider literature \cite{balesni_2024_towards,clymer_2024_safety,buhl_2025_an}, could form important parts of safety evaluation of frontier AI more generally. We aim to ameliorate those issues here. 

\subsection{Risk and Hazard Identification}

The typical way in which risk management is constructed within safety engineering involves identifying hazards and hazardous events, assessing, controlling and monitoring the risk of these hazards, setting out an acceptable level of risk, documenting those risks formally, and placing this information into a safety argument which contextualises and justifies the risk management process and outputs. It is an iterative and through-life process. Safety management within safety assurance always begins with scoping the system and its context and identifying \textit{hazards }or \textit{hazardous events} and \textit{risks} \cite{internationalstandardsorganisation_2014_isoiec}.
\begin{itemize}
    \item \textit{Hazardous event: }An event that can cause harm 
    \item \textit{Hazard}: Potential source of harm
    \item \textit{Risk: }The combination of the probability of occurrence of harm and the severity of that harm
\end{itemize}

\subsubsection{Hazards}

There are various approaches to hazard analysis. Hazards must be approached consistently and defined clearly, which is a key consideration for frontier AI. The identification of hazards or hazardous events underpins safety assurance. There are new kinds of hazardous events presented by increasingly advanced AI. The focus in the alignment safety case literature is on catastrophically dangerous misalignment \cite{clymer_2024_safety,balesni_2024_towards}, as well as associated hazards, such as those posed by CBRN capabilities \cite{buhl_2024_safety}. Both approaches focus squarely on the capabilities presented by frontier AI systems, hence the focus on novel hazardous capabilities. Hazardous events in ISO/IEC Guide 51:2014 appear to be analogous to the term “threat model” used in the alignment literature for risks posed by frontier systems themselves \cite{clymer_2024_safety}.

Nonetheless, model developers will have to identify which hazards or hazardous events are of concern. It is unclear currently which hazardous events should be identified in a frontier AI system \cite{hilton_2025_safety}. Within safety assurance, hazards are those which could cause physical, environmental or increasingly some psychological harm \cite{FearnleyForthcoming-FEACCI-2}. Moreover, safety assurance has grappled with the expanding scope of harms in safety-critical systems in recent years, towards sociotechnical understandings of how systems are developed \cite{dobbe_2022_system}. This shift has been marked explicitly by the introduction of AI-based systems into safety-critical settings \cite{burr_2022_ethical}. It has therefore been accompanied by considerable debate across the field, building on an extensive body of academic, policy and industry best practice. The question of who decides on which type of hazards should be included is a pertinent question for any suggestion that a regulator will evaluate safety cases \cite{davies_2023_regulating}. 

The question of hazards and harms is perhaps more complex than it may initially seem. For example, \citeauthor{clymer_2024_safety} suggest that any catastrophic harm is relevant. Other research is motivated by catastrophe caused by scheming or potentially superintelligent models \cite{korbak_2025_how}. \citeauthor{clymer_2024_safety} define catastrophic harm as ‘large-scale devastation of a specific severity’. Their definition involves “billions of dollars in damages or thousands of deaths” \cite{clymer_2024_safety}. \citeauthor{buhl_2024_safety} consider a broader scope of hazards, such as ability for the model to assist with weapon creation \cite{buhl_2024_safety}.

Misalignment could ostensibly cause many harms outside of catastrophic or existential harm. For example, sycophancy could amplify bias or cause certain psychological harms \cite{sharma_2023_towards}. However, some definitions omit these harms in order to focus their safety analysis onto specific hazards \cite{clymer_2024_safety}. Similarly, frontier developers have explicitly noted that their duties are not to push the frontier of capabilities without suitable risk management processes \cite{anthropic_2024_responsible,googledeepmind_2025_frontier,openai_2025_gpt5}.

However, there are potentially some inconsistencies in the current approach to hazard analysis. For example, it is well-established that goal misspecification is a hazard \cite{malekmechergui_2024_goal}. If goal misspecification were present in some cases but not others, does it only become a hazardous event once it reaches the threshold of causing billions of dollars of damages? Or when a certain amount of deaths could be reasonably foreseen? Or is it only a hazard where the system is also superintelligent and is intentionally scheming? If so, who decides? This is where risk management techniques and associated goal-based policies begin to play an important aspect of safety argumentation.

\subsubsection{Risk}

Within safety management, the relevant and affected stakeholders, e.g. regulators, users and developers, need to decide on the amount of residual risk - necessarily existing risk - which they are willing to tolerate. For example, within nuclear safety, there is always a risk, however minimal, of a radiological spill, as this is core to the functioning of the plant.

\paragraph{Hazard Elimination}

The first line of defence within risk management is to eliminate hazards. However, hazard elimination may not always be possible, particularly given the complexity of frontier AI development, requiring risk reduction. For example, RLHF aims to align models with human preferences, but there remains a risk of jailbreaks or harmful output \cite{casper_2023_open}.

\paragraph{Risk reduction}

There may be some hazards which developers are unable to remove. If you are unable to remove a hazard, you aim to reduce the risk of the hazard (by targeting likelihood and/or severity). For example, in a nuclear plant, you may introduce protective equipment for those interacting with a harmful chemical substance which cannot be removed. Similarly, within frontier model development, you may aim to introduce guardrails on a model at post-training or deployment stage.  For example, OpenAI and Apollo Research recently presented an implicit risk reduction argument that they can substantially reduce the risk of scheming via deliberative alignment, from over 10\% to under 1\% \cite{schoen_2025_stress}. The remaining risk of scheming which is not covered by the technique may fall under the residual risk accepted by the developer.

There are various approaches to analysing and reducing the risk of hazards in the literature. Risk reduction can take on various forms. The framing of risk reduction can be broken down into two parts:
\begin{enumerate}
    \item \textbf{\textbf{Modify the design or operating procedure within model development or deployment: }}For example, what decisions were made during pre-training or post-training to reduce the likelihood of certain hazardous events occurring? 
    \item \textbf{\textbf{Reduce the severity of consequences: }} For example, mitigate who can use the model and place guardrails to limit the model’s propensity to carry out harmful behaviours.
\end{enumerate}

\subsection{Risk reduction Methods}

\textbf{Qualitative assessments:}  The vast majority of alignment safety case analyses are qualitative, engaging in debate or reasoning about the likelihood of misalignment. Outside of the alignment safety case literature, authors who have written on frontier AI safety and considered safety-critical systems also note the utility of quantitative assessments \cite{dalrymple_2024_towards}.

Qualitative arguments are therefore an accepted method in both the safety assurance and alignment communities \cite{graydon_2017_an}. As a method for risk reduction, reasoning through the propensity of certain hazards or threat models, and considering how those might be evaluated or red-teamed, can be a method for reducing and assessing risk. For example, \citeauthor{shlegeris_2025_ai} notes a variety of ways in which one can analyse or consider the propensity of a model to attempt to escape \cite{shlegeris_2025_ai}. \citeauthor{sharkey_2025_open} discuss how a safety case might be informed by a qualitative understanding of the internal characteristics of a model, derived from mechanistic interpretability research \cite{sharkey_2025_open}. These approaches are familiar and would implicitly be endorsed by those working on safety across disciplines.

\textbf{Safety Integrity Levels: } Another method through which risk reduction can be achieved in safety-critical systems is to specify certain Safety Integrity Levels (SILs) which must be reached \cite{redmill_1999_understanding}. An SIL is determined firstly by a risk assessment, which then outlines the target SIL required, before considering what the relevant risk reduction factor must be and how the safety around that risk reduction is achieved.

There is already a potential analogue in frontier AI safety: security levels, such as Critical Capability Levels \cite{googledeepmind_2025_frontier} or AI Safety Levels \cite{anthropic_2024_responsible}, set by frontier developers in response to voluntary commitments. SILs tend to be underpinned by industry standards. Frontier AI security levels outline certain capabilities which must be safeguarded against under certain voluntary commitments. Furthermore, there is active research into standardisation within frontier AI \cite{ziosi_2025_safety,roberts_2025_can}. It would be useful to have particular industry standards for the relevant subcomponents - or hazards - which form part of these security levels. For example, within automotive, companies can follow the Automotive Safety Integrity Level (ASILs) as defined in the international automotive standard ISO 26262 \cite{internationalstandardsorganisation_2018_iso2626292018}. Despite long-standing reservations about SILs \cite{mcdermid_2001_scs}, presenting equivalent standards in frontier AI for safety cases could be an impactful research direction, helping to clarify the safety process and metrics followed by frontier AI developers.

\textbf{Deterministic methods and method transferability:} It seems likely that many quantitative methods from safety assurance may struggle to operate within the non-deterministic, highly uncertain field of frontier AI. For example, the field of formal methods will likely struggle to cope with the complexity of DNNs with general-purpose capabilities, given long-standing concerns about their utility and practicality for more modest software systems \cite{littlewood_1993_validation}. In contrast, some theories from systems safety, such as emergence \cite{lintern_2022_emergence}, may offer interesting insights to those grappling with emergent capabilities of LLMs \cite{wei_2022_emergent}. Others have pointed to the fact that early safety analysis of advanced AI systems, particularly early reinforcement learning systems, is framed explicitly within the approach of hazard analysis \cite{harding_2025_what}.

As a result, we believe that there will be relevant techniques that can be transferred from safety assurance to frontier AI and consequently applied to a risk assessment.

\subsubsection{Residual risks}

Once risks have been considered and analysed, there are some risks which will always be residual and cannot be fully designed out. This is particularly pertinent to the context of LLMs, which present a range of risks of uncertain quantity or harm. Safety cases for safety-critical systems would require developers to document all actions taken to resolve these residual risks. The ‘risk owner’ or duty holder is responsible for deciding whether to accept the risk or apply additional resources \cite{healthandsafetyexecutive_2001_reducing}. This is implicit in many of the actions taken by frontier AI developers, such as safety documentation \cite{anthropic_2024_responsible,googledeepmind_2025_frontier}, system cards \cite{anthropic_2025_system,openai_2025_gpt5,googledeepmind_2025_gemini} or model releases which trigger new security levels.

Residual risk analysis is feasible for frontier AI systems. An example risk reduction analysis of CBRN capabilities at a basic level might involve breaking potential risks into various steps:

\begin{enumerate}
    \item \textbf{Analysis: }CBRN capability is a result of pre-training data including information about CBRN-relevant topics \cite{chen_2025_enhancing}. This cannot be completely designed out, due to the scale of pre-training data.
    \item \textbf{Proposed Solution: }Given (1), one method might be RLHF (Reinforcement Learning from Human Feedback). However, this method is imperfect \cite{casper_2023_open}, so further guardrails may be required to reduce the risk further. 
    \item \textbf{Residual Risk Management: }Introduce deliberative alignment \cite{guan_2024_deliberative} to reduce the residual risk of harmful requests still outputting harmful content despite RLHF.  
\end{enumerate}

The risk owner may then choose to “own” or accept the risk of increased sycophancy due to RLHF \cite{casper_2023_open}.

The benefit of this explicit process of risk reduction has a dual function: it provides validation and evidence to system developers that they have considered risks as systematically as possible; and it documents to others that they took relevant actions at correct stages by presenting processes and arguments supported by their body of evidence. We present an illustrative example of the risk reduction workflow for a scheming model in Appendix C, in line with the relevant ISO-IEC 51-2014 standard \cite{internationalstandardsorganisation_2014_isoiec}.

\subsection{Risk reduction tools}

There are various risk reduction tools available within safety assurance. Some of these tools are alluded to by those in the alignment safety case literature. We explore briefly some relevant overlaps between these tools and approaches by those in frontier AI safety.

\subsubsection{Risk Assessment}

The Risk Assessment within the safety assurance community tends to involve a quantum of probability and severity of a hazard occurring. In a safety-critical risk assessment, if there were even a negligible chance of a model deployment causing existential catastrophe or total disempowerment \cite{clymer_2024_safety}, the severity quantum would appear to be infinite, given the scale of the harm.

This indicates the need for rigorous, context-specific risk assessment. This conundrum is ameliorated by using the “ALARP” principle - “as low as reasonably practicable” - which underpins safety assurance regulation in the U.K. \cite{haddoncavekc_the,healthandsafetyexecutive_2001_reducing}. Where a risk owner considers that it may be grossly disproportionate to the improvement gained to continue to limit certain risks, they might choose to deploy a system regardless \cite{haddoncavekc_the}. 

Risk assessments could take on various forms. When presenting system cards or certain pieces of research, model developers implicitly present risk mitigations, such as evidence that a model will not output CBRN-relevant content \cite{openai_2025_gptoss120b}.

The aim of the risk assessment is to build an argument internally and externally that your system will be safe. We argue that this could be a promising assurance method within frontier AI. Despite potential differences in method, a risk assessment in both the alignment safety case and safety assurance communities does need to take place in order to understand the threats posed by certain model capabilities and behaviours, and thereafter to make an argument about the safety of the model.

\subsubsection{Management tools such as Hazard Logs}

Hazard logs are safety management tools which allow organisations to keep track of which hazards must be mitigated. A hazard tracking system may also link to a hazard log, in order to allocate actions to reduce the risk for each unacceptable hazard \cite{ministryofdefence_2025_hazard}. 

In some ways, the frontier AI community is already well-placed, with a rich body of associated literature, to engage in hazard logging. Model developers publish and engage in evals \cite{fronsdal_2025_petri}, blog on safety concerns \cite{kalai_2025_why, anthropic_2024_three} and publish system cards outlining potential risks and how they were mitigated \cite{anthropic_2025_system,openai_2025_gpt5,googledeepmind_2025_gemini}.

\subsubsection{Setting derived safety requirements}

Among the most significant challenges presenting safety cases for advanced AI is the setting of derived safety requirements. Within other fields, we have seen that there are well-established regulations and standards which set out industry-set obligations, which lead to derived safety requirements from those obligations. Secondly, even if a safety case were not required by statute, standards bodies and industry organisations can guide best practice. This is presently not the case within frontier AI. However, we note the importance of research directions aiming to ameliorate this concern, some of which are ongoing \cite{simpson_2025_voluntary}.

\section{GSN-based Safety Argument}
We present a GSN-based case study (Figure 1) which aims to show how a safety argument may be supported by relevant risk-related evidence to present a clear argument, which can be interrogated to increase confidence in the risk assessment and hazard analysis process. We present two example hazardous events: Deceptive Alignment and CBRN capabilities. \citeauthor{hilton_2025_safety} recognise that there are various open problems in the construction of a safety case for frontier AI systems, such as which notation to use or which top-level goal to present \cite{hilton_2025_safety}.

GSN is a widely used notation for capturing safety case arguments in high-risk industries, with 2014 evidence indicating that 3/4 of all UK military aircraft used a GSN-based safety case \cite{researchexcellenceframework_2014_com04} and NASA  publishing GSN-based studies in their research repository \cite{denney_2012_hierarchical,witulski_2016_goal}. We place the full argument in Appendix A. The aim of developing the safety argument is to help those building it to consider the steps they have taken to assure the system and consequently identify issues or gaps in that process.

The GSN (Figure 1), covered in full detail in Appendix A, begins at a top-level goal (Frontier AI System does not lead to catastrophic impact). The top-level goal is supported by examination of identified hazardous events (CBRN capabilities and Deceptive Alignment). Each hazardous event is accompanied by supporting goals and evidence, which could be derived from an earlier risk assessment, risk reduction or evaluation methods, or from derived safety requirements from existing standards. Evidence in GSN, i.e. solutions, are not claims but references to data or results. 

Control of the hazardous event is supported by arguments over through-life controls and mitigations, split here into development, deployment and post-deployment. This highlights a holistic consideration of model development, from pre-training analysis through to post-deployment monitoring. Each goal is then supported by evidence. 

These arguments combine to present a structured, auditable trail of the steps taken to achieve the top-level goal by providing evidence that certain hazardous events have been controlled as thoroughly as possible for deployment. We aim to build on this foundation in later work. 

\begin{sidewaysfigure}[p]
    \centering
    \includegraphics[width=\textwidth, height=\textheight, keepaspectratio]{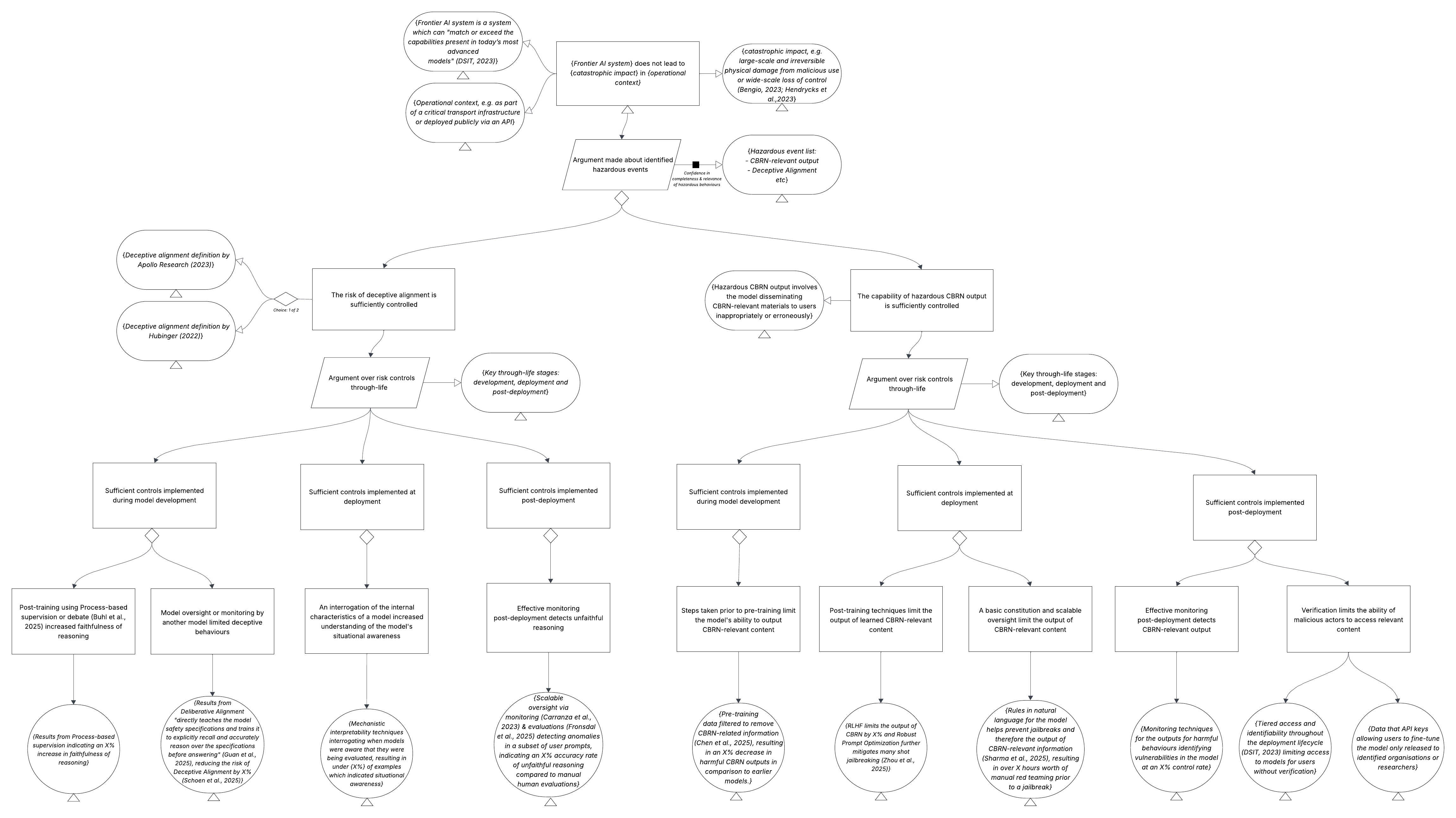}
    \caption{The full GSN can be traced from a top-level goal down to supporting evidence for relevant subgoals and strategies. The evidence is drawn from existing research and solutions within frontier AI safety literature \cite{departmentforscienceinnovationandtechnology_2023_capabilities,hendrycks_2023_an,buhl_2025_an,guan_2024_deliberative,schoen_2025_stress,carranza_2023_deceptive,fronsdal_2025_petri,chen_2025_enhancing,sharma_2025_constitutional,apolloresearch_2023_understanding,hubinger_2022_how,zhou_2024_robust}.}
    \label{Full GSN}
\end{sidewaysfigure}

\subsection{Future research}

\textbf{Governance infrastructure: }Governance infrastructure underpins safety cases across safety-critical systems. Within automotive, developers have ISO26262 \cite{internationalstandardsorganisation_2018_iso2626292018}. Within energy, developers are required to produce safety cases under The Offshore Installations (Safety Case) Regulations 1992. Existing voluntary commitments and a growing body of policy-based work which interacts with frontier AI safety frameworks may help to ameliorate the lack of relevant standards in frontier AI safety \cite{ziosi_2025_safety,simpson_2025_voluntary}. We present this as a significant area for future research.

\textbf{Argument patterns for LLMs: }There is a rich body of literature on argument patterns for safety-critical industries. Furthermore, there is a growing body of literature within alignment safety cases on using certain patterns to assure LLMs. We hope to build on the work presented in this paper in future work. Indeed, this work could build on cross-disciplinary frameworks presented elsewhere in the assurance literature \cite{burr_2022_ethical,habli_2025_the,barrett_2025_assessing}.

\section{Conclusion}

The alignment safety case literature has been immensely influential and it has provided a valuable way of thinking about deployment-related risks. However, risk management and ensuing safety cases require careful, through-life consideration around system capabilities. If safety cases are to make up core components of frontier AI companies’ risk frameworks and be a global research priority, these foundations need to be robust. Our work has aimed to set a new foundation for frontier AI safety cases, recentring the work in best practice in safety assurance and novel alignment techniques.

\medskip

\clearpage
\begin{ack}
    This work was supported by UKRI AI Centre for Doctoral Training in Safe Artificial Intelligence Systems (SAINTS) (EP/Y030540/1).
\end{ack}
\bibliography{IASEAI}

@misc{bengio_2025_the,
  author = {Bengio, Yoshua and Maharaj, Tegan and Ong, Luke and Russell, Stuart and Song, Dawn and Tegmark, Max and Xue, Lan and Zhang, Ya-Qin and Casper, Stephen and Lee, Wan Sie and Mindermann, Sören and Wilfred, Vanessa and Balachandran, Vidhisha and Barez, Fazl and Belinsky, Michael and Bello, Imane and Bourgon, Malo and Brakel, Mark and Campos, Siméon and Cass-Beggs, Duncan and Chen, Jiahao and Chowdhury, Rumman and Seah, Kuan Chua and Clune, Jeff and Dai, Juntao and Delaborde, Agnes and Dziri, Nouha and Eiras, Francisco and Engels, Joshua and Fan, Jinyu and Gleave, Adam and Goodman, Noah and Heide, Fynn and Heidecke, Johannes and Hendrycks, Dan and Hodes, Cyrus and Low, Bryan and Huang, Minlie and Jawhar, Sami and Jingyu, Wang and Kalai, Adam Tauman and Kamphuis, Meindert and Kankanhalli, Mohan and Kantamneni, Subhash and Kirk, Mathias Bonde and Kwa, Thomas and Ladish, Jeffrey and Lam, Kwok-Yan and Sie, Wan Lee and Lee, Taewhi and Li, Xiaojian and Liu, Jiajun and Lu, Chaochao and Mai, Yifan and Mallah, Richard and Michael, Julian and Moës, Nick and Möller, Simon and Nam, Kihyuk and Ng, Kwan Yee and Nitzberg, Mark and Nushi, Besmira and hÉigeartaigh, Seán O and Ortega, Alejandro and Peigné, Pierre and Petrie, James and Prud'Homme, Benjamin and Rabbany, Reihaneh and Sanchez-Pi, Nayat and Schwettmann, Sarah and Shlegeris, Buck and Siddiqui, Saad and Sinha, Aradhana and Soto, Martín and Tan, Cheston and Ting, Dong and Tjhi, William and Trager, Robert and Tse, Brian and Tung, Anthony and Wilfred, Vanessa and Willes, John and Wong, Denise and Xu, Wei and Xu, Rongwu and Zeng, Yi and Zhang, HongJiang and Žikelić, Djordje},
  title = {The Singapore Consensus on Global AI Safety Research Priorities},
  url = {https://arxiv.org/abs/2506.20702},
  urldate = {2025-10-10},
  year = {2025},
  organization = {arXiv.org}
}

@misc{bengio_2025_international,
  author = {Bengio, Yoshua and Mindermann, Sören and Privitera, Daniel and Besiroglu, Tamay and Bommasani, Rishi and Casper, Stephen and Choi, Yejin and Fox, Philip and Garfinkel, Ben and Goldfarb, Danielle and Heidari, Hoda and Ho, Anson and Kapoor, Sayash and Khalatbari, Leila and Longpre, Shayne and Manning, Sam and Mavroudis, Vasilios and Mazeika, Mantas and Michael, Julian and Newman, Jessica and Ng, Kwan Yee and Okolo, Chinasa T and Raji, Deborah and Sastry, Girish and Seger, Elizabeth and Skeadas, Theodora and South, Tobin and Strubell, Emma and Tramèr, Florian and Velasco, Lucia and Wheeler, Nicole and Acemoglu, Daron and Adekanmbi, Olubayo and Dalrymple, David and Dietterich, Thomas G and Felten, Edward W and Fung, Pascale and Gourinchas, Pierre-Olivier and Heintz, Fredrik and Hinton, Geoffrey and Jennings, Nick and Krause, Andreas and Leavy, Susan and Liang, Percy and Ludermir, Teresa and Marda, Vidushi and Margetts, Helen and McDermid, John and Munga, Jane and Narayanan, Arvind and Nelson, Alondra and Neppel, Clara and Oh, Alice and Ramchurn, Gopal and Russell, Stuart and Schaake, Marietje and Schölkopf, Bernhard and Song, Dawn and Soto, Alvaro and Tiedrich, Lee and Varoquaux, Gaël and Yao, Andrew and Zhang, Ya-Qin and Albalawi, Fahad and Alserkal, Marwan and Ajala, Olubunmi and Avrin, Guillaume and Busch, Christian and Ponce, Carlos and Fox, Bronwyn and Gill, Amandeep Singh and Hatip, Ahmet Halit and Heikkilä, Juha and Jolly, Gill and Katzir, Ziv and Kitano, Hiroaki and Krüger, Antonio and Johnson, Chris and Khan, Saif M and Lee, Kyoung Mu and Ligot, Dominic Vincent and Molchanovskyi, Oleksii and Monti, Andrea and Mwamanzi, Nusu and Nemer, Mona and Oliver, Nuria and Portillo, José Ramón López  and Ravindran, Balaraman and Rivera, Raquel Pezoa and Riza, Hammam and Rugege, Crystal and Seoighe, Ciarán and Sheehan, Jerry and Sheikh, Haroon and Wong, Denise and Zeng, Yi},
  title = {International AI Safety Report},
  url = {https://arxiv.org/abs/2501.17805},
  year = {2025},
  organization = {arXiv.org}
}

@misc{ministryofdefence_2007_defence,
  author = {Ministry of Defence},
  publisher = {Glasgow: UK Defence Standardization},
  title = {Defence Standard 00-56 Safety Management Requirements for Defence Systems.},
  year = {2007}
}

@inproceedings{knight_2002_safety,
  author = {Knight, John C.},
  title = {Safety critical systems},
  doi = {10.1145/581339.581406},
  year = {2002},
  booktitle = {Proceedings of the 24th international conference on Software engineering }
}

@article{burton_2019_mind,
  author = {Burton, Simon and Habli, Ibrahim and Lawton, Tom and McDermid, John and Morgan, Phillip and Porter, Zoe},
  month = {11},
  pages = {103201},
  title = {Mind the Gaps: Assuring the Safety of Autonomous Systems from an Engineering, Ethical, and Legal Perspective},
  doi = {10.1016/j.artint.2019.103201},
  volume = {279},
  year = {2019},
  journal = {Artificial Intelligence}
}

@inproceedings{sandom_2002_human,
  author = {Sandom, Carl},
  pages = {125-139},
  publisher = {Springer},
  title = {Human Factors Considerations for System Safety},
  year = {2002},
  booktitle = {Proceedings of the Tenth Safety-critical Systems Symposium}
}

@article{rae_2020_a,
  author = {Rae, Andrew and Provan, David and Aboelssaad, Hossam and Alexander, Rob},
  month = {06},
  title = {A manifesto for Reality-based Safety Science},
  doi = {10.1016/j.ssci.2020.104654},
  urldate = {2020-10-30},
  volume = {126},
  year = {2020},
  journal = {Safety Science}
}

@misc{graydon_2017_the,
  author = {Graydon, Patrick},
  title = {The Safety Argumentation Schools of Thought},
  url = {https://shemesh.larc.nasa.gov/people/msg/graydon2017thesasots.pdf},
  year = {2017}
}

@misc{habli_2025_the,
  author = {Habli, Ibrahim and Hawkins, Richard and Paterson, Colin and Ryan, Philippa and Jia, Yan and Sujan, Mark and McDermid, John},
  title = {The BIG Argument for AI Safety Cases},
  url = {https://arxiv.org/abs/2503.11705},
  year = {2025},
  organization = {arXiv.org}
}

@misc{witulski_2016_goal,
  author = {Witulski, Arthur and Austin, Rebekah and Evans, John and Mahadevan, Nag and Karsai, Gabor and Sierawski, Brian and Label, Ken and Reed, Robert and Schrimpf, Ron},
  publisher = {NASA Electronic Parts and Packing Program},
  title = {Goal Structuring Notation in a Radiation Hardening Assurance Case for COTS-Based Spacecraft},
  url = {https://ntrs.nasa.gov/api/citations/20160007995/downloads/20160007995.pdf},
  urldate = {2025-10-10},
  year = {2016},
  organization = {nepp.nasa.gov}
}

@misc{denney_2012_hierarchical,
  author = {Denney, Ewen and Whiteside, Iain},
  publisher = {NASA STI Repository},
  title = {Hierarchical Safety Cases},
  url = {https://ntrs.nasa.gov/api/citations/20130001737/downloads/20130001737.pdf},
  urldate = {2025-10-10},
  year = {2012},
  organization = {NASA STI Repository}
}

@phdthesis{kelly_1999_arguing,
  author = {Kelly, Timothy},
  title = {Arguing Safety -- A Systematic Approach to Managing Safety Cases Timothy Patrick Kelly},
  year = {1999},
  school = {University of York}
}

@misc{denney_2017_dynamic,
  author = {Denney, Ewen and Pai, Ganesh and Habli, Ibrahim},
  publisher = {NASA STI Repository},
  title = {Dynamic Safety Cases for Through-life Safety Assurance},
  url = {https://ntrs.nasa.gov/api/citations/20150011054/downloads/20150011054.pdf},
  year = {2017}
}

@inproceedings{palin_2010_assurance,
  author = {Palin, Robert  and Habli, Ibrahim },
  pages = {82-96},
  publisher = {Springer},
  title = {Assurance of Automotive Safety – A Safety Case Approach},
  volume = {6351},
  year = {2010},
  organization = {Lecture Notes in Computer Science},
  booktitle = {SAFECOMP'10: Proceedings of the 29th international conference on Computer safety, reliability, and security}
}

@misc{thehealthfoundation_2023_using,
  author = {The Health Foundation},
  title = {Using safety cases in industry and healthcare},
  url = {https://www.health.org.uk/reports-and-analysis/reports/using-safety-cases-in-industry-and-healthcare},
  year = {2023},
  organization = {The Health Foundation}
}

@misc{healthandsafetyexecutive_2024_preparing,
  author = {Health and Safety Executive},
  publisher = {U.K. Government},
  title = {Preparing a Safety Case Report},
  url = {https://www.gov.uk/guidance/preparing-a-safety-case-report},
  year = {2024}
}

@misc{anthropic_2024_responsible,
  author = {Anthropic},
  title = {Responsible Scaling Policy},
  url = {https://www-cdn.anthropic.com/872c653b2d0501d6ab44cf87f43e1dc4853e4d37.pdf},
  year = {2024}
}

@misc{googledeepmind_2025_frontier,
  author = {Google DeepMind},
  month = {02},
  title = {Frontier Safety Framework 2.0},
  url = {https://storage.googleapis.com/deepmind-media/DeepMind.com/Blog/updating-the-frontier-safety-framework/Frontier%20Safety%20Framework%202.0%20(1).pdf},
  urldate = {2025-10-10},
  year = {2025}
}

@article{rae_2017_probative,
  author = {Rae, Andrew and Alexander, Rob D.},
  month = {02},
  pages = {190-204},
  title = {Probative blindness and false assurance about safety},
  doi = {10.1016/j.ssci.2016.10.005},
  urldate = {2020-02-08},
  volume = {92},
  year = {2017},
  journal = {Safety Science}
}

@book{downer_2024_rational,
  author = {Downer, John},
  publisher = {The MIT Press},
  title = {Rational Accidents},
  doi = {10.7551/mitpress/8844.001.0001},
  urldate = {2024-10-17},
  year = {2024},
  organization = {The MIT Press eBooks}
}

@misc{nuclearenergyagency_2004_management,
  author = {Nuclear Energy Agency and Organisation for Economic Co-Operation and Development},
  publisher = {Workshop Proccedings Stockholm, Sweden},
  title = {Management of Uncertainty in Safety Cases and the Role of Risk },
  url = {https://www.oecd-nea.org/upload/docs/application/pdf/2020-12/nea5302-management-uncertainty-risk.pdf},
  year = {2004}
}

@misc{bowen_2025_ai,
  author = {Bowen, Dillon and Dombrowski, Ann-Kathrin and Gleave, Adam and Cundy, Chris},
  title = {AI Companies Should Report Pre- and Post-Mitigation Safety Evaluations},
  url = {https://arxiv.org/abs/2503.17388v1},
  urldate = {2025-10-10},
  year = {2025},
  organization = {arXiv.org}
}

@misc{defencesafetyauthority_2024_dsa,
  author = {Defence Safety Authority},
  publisher = {U.K. Government},
  title = {DSA 03.OME Part 1: Defence Code of Practice (DCOP) 113: OME Through-Life Capability Management (TLCM)},
  url = {https://assets.publishing.service.gov.uk/media/689f155d2e8cc8ec5b3572fd/DSA_03.OME_Part_1_DCOP_113_-_OME_Through_Life_Capability_Management_-_TLCM.pdf},
  year = {2024}
}

@misc{greenblatt_2025_focus,
  author = {Greenblatt, Ryan},
  month = {09},
  publisher = {Redwood Research blog},
  title = {Focus transparency on risk reports, not safety cases},
  url = {https://blog.redwoodresearch.org/p/focus-transparency-on-risk-reports?hide_intro_popup=true},
  urldate = {2025-10-10},
  year = {2025},
  organization = {Redwoodresearch.org}
}

@inproceedings{dobbe_2022_system,
  author = {Dobbe, Roel},
  month = {06},
  title = {System Safety and Artificial Intelligence},
  doi = {10.1145/3531146.3533215},
  urldate = {2022-10-21},
  year = {2022},
  booktitle = {FAccT '22: Proceedings of the 2022 ACM Conference on Fairness, Accountability, and Transparency}
}

@article{burr_2022_ethical,
  author = {Burr, Christopher and Leslie, David},
  month = {06},
  title = {Ethical assurance: a practical approach to the responsible design, development, and deployment of data-driven technologies},
  doi = {10.1007/s43681-022-00178-0},
  volume = {3},
  year = {2022},
  journal = {AI and Ethics}
}

@misc{openai_2025_gptoss120b,
  author = {OpenAI and Agarwal, Sandhini and Ahmad, Lama and Ai, Jason and Altman, Sam and Applebaum, Andy and Arbus, Edwin and Arora, Rahul K and Bai, Yu and Baker, Bowen and Bao, Haiming and Barak, Boaz and Bennett, Ally and Bertao, Tyler and Brett, Nivedita and Brevdo, Eugene and Brockman, Greg and Bubeck, Sebastien and Chang, Che and Chen, Kai and Chen, Mark and Cheung, Enoch and Clark, Aidan and Cook, Dan and Dukhan, Marat and Dvorak, Casey and Fives, Kevin and Fomenko, Vlad and Garipov, Timur and Georgiev, Kristian and Glaese, Mia and Gogineni, Tarun and Goucher, Adam and Gross, Lukas and Guzman, Katia Gil and Hallman, John and Hehir, Jackie and Heidecke, Johannes and Helyar, Alec and Hu, Haitang and Huet, Romain and Huh, Jacob and Jain, Saachi and Johnson, Zach and Koch, Chris and Kofman, Irina and Kundel, Dominik and Kwon, Jason and Kyrylov, Volodymyr and Le, Elaine Ya and Leclerc, Guillaume and Lennon, James Park and Lessans, Scott and Lezcano-Casado, Mario and Li, Yuanzhi and Li, Zhuohan and Lin, Ji and Liss, Jordan and Liu, Lily and Liu, Jiancheng and Lu, Kevin and Lu, Chris and Martinovic, Zoran and McCallum, Lindsay and McGrath, Josh and McKinney, Scott and McLaughlin, Aidan and Mei, Song and Mostovoy, Steve and Mu, Tong and Myles, Gideon and Neitz, Alexander and Nichol, Alex and Pachocki, Jakub and Paino, Alex and Palmie, Dana and Pantuliano, Ashley and Parascandolo, Giambattista and Park, Jongsoo and Pathak, Leher and Paz, Carolina and Peran, Ludovic and Pimenov, Dmitry and Pokrass, Michelle and Proehl, Elizabeth and Qiu, Huida and Raila, Gaby and Raso, Filippo and Ren, Hongyu and Richardson, Kimmy and Robinson, David and Rotsted, Bob and Salman, Hadi and Sanjeev, Suvansh and Schwarzer, Max and Sculley, D and Sikchi, Harshit and Simon, Kendal and Singhal, Karan and Song, Yang and Stuckey, Dane and Sun, Zhiqing and Tillet, Philippe and Toizer, Sam and Tsimpourlas, Foivos and Vyas, Nikhil and Wallace, Eric and Wang, Xin and Wang, Miles and Watkins, Olivia and Weil, Kevin and Wendling, Amy and Whinnery, Kevin and Whitney, Cedric and Wong, Hannah and Yang, Lin and Yang, Yu and Yasunaga, Michihiro and Ying, Kristen and Zaremba, Wojciech and Zhan, Wenting and Zhang, Cyril and Zhang, Brian and Zhang, Eddie and Zhao, Shengjia},
  title = {gpt-oss-120b \& gpt-oss-20b Model Card},
  url = {https://arxiv.org/abs/2508.10925},
  year = {2025},
  organization = {arXiv.org}
}

@inproceedings{malekmechergui_2024_goal,
  author = {Malek Mechergui and Sarath Sreedharan},
  month = {03},
  pages = {10110-10118},
  publisher = {Association for the Advancement of Artificial Intelligence},
  title = {Goal Alignment: Re-analyzing Value Alignment Problems Using Human-Aware AI},
  doi = {10.1609/aaai.v38i9.28875},
  url = {https://ojs.aaai.org/index.php/AAAI/article/view/28875},
  volume = {38},
  year = {2024},
  booktitle = {Proceedings of the AAAI Conference on Artificial Intelligence}
}

@misc{schoen_2025_stress,
  author = {Schoen, Bronson and Nitishinskaya, Evgenia and Balesni, Mikita and Højmark, Axel and Hofstätter, Felix and Scheurer, Jérémy and Meinke, Alexander and Wolfe, Jason and van der Weij, Teun and Lloyd, Alex and Goldowsky-Dill, Nicholas and Fan, Angela and Matveiakin, Andrei and Shah, Rusheb and Williams, Marcus and Glaese, Amelia and Barak, Boaz and Zaremba, Wojciech and Hobbhahn, Marius},
  title = {Stress Testing Deliberative Alignment for Anti-Scheming Training},
  url = {https://www.arxiv.org/abs/2509.15541},
  year = {2025},
  organization = {arXiv.org}
}

@misc{dalrymple_2024_towards,
  author = {Dalrymple, David davidad and Skalse, Joar and Bengio, Yoshua and Russell, Stuart and Tegmark, Max and Seshia, Sanjit and Omohundro, Steve and Szegedy, Christian and Goldhaber, Ben and Ammann, Nora and Abate, Alessandro and Halpern, Joe and Barrett, Clark and Zhao, Ding and Zhi-Xuan, Tan and Wing, Jeannette and Tenenbaum, Joshua},
  title = {Towards Guaranteed Safe AI: A Framework for Ensuring Robust and Reliable AI Systems},
  url = {https://arxiv.org/abs/2405.06624},
  urldate = {2024-10-14},
  year = {2024},
  organization = {arXiv.org}
}

@article{graydon_2017_an,
  author = {Graydon, Patrick J and C. Michael Holloway},
  month = {02},
  pages = {53-65},
  title = {An investigation of proposed techniques for quantifying confidence in assurance arguments},
  doi = {10.1016/j.ssci.2016.09.014},
  urldate = {2023-05-16},
  volume = {92},
  year = {2017},
  journal = {Safety Science}
}

@misc{shlegeris_2025_ai,
  author = {Shlegeris, Buck},
  title = {AI Control and Why AI Security People Should Care},
  url = {https://www.far.ai/events/sessions/buck-shlegeris-ai-control-and-why-ai-security-people-should-care},
  year = {2025}
}

@misc{sharkey_2025_open,
  author = {Sharkey, Lee and Chughtai, Bilal and Batson, Joshua and Lindsey, Jack and Wu, Jeff and Bushnaq, Lucius and Goldowsky-Dill, Nicholas and Heimersheim, Stefan and Ortega, Alejandro and Bloom, Joseph and Biderman, Stella and Garriga-Alonso, Adria and Conmy, Arthur and Nanda, Neel and Rumbelow, Jessica and Wattenberg, Martin and Schoots, Nandi and Miller, Joseph and Michaud, Eric J and Casper, Stephen and Tegmark, Max and Saunders, William and Bau, David and Todd, Eric and Geiger, Atticus and Geva, Mor and Hoogland, Jesse and Murfet, Daniel and McGrath, Tom},
  title = {Open Problems in Mechanistic Interpretability},
  url = {https://arxiv.org/abs/2501.16496},
  urldate = {2025-03-05},
  year = {2025},
  organization = {arXiv.org}
}

@misc{internationalstandardsorganisation_2018_iso2626292018,
  author = {International Standards Organisation},
  title = {ISO26262-9:2018},
  url = {https://www.iso.org/standard/68391.html},
  year = {2018}
}

@misc{roberts_2025_can,
  author = {Roberts, Huw and Ziosi, Marta},
  publisher = {Oxford Martin AI Governance Initiative},
  title = {Can we standardise the frontier of AI?},
  url = {https://aigi.ox.ac.uk/wp-content/uploads/2025/06/ssrn-5271446.pdf},
  urldate = {2025-10-10},
  year = {2025}
}

@misc{ziosi_2025_safety,
  author = {Ziosi, Marta and Gealy, James and Plueckebaum, Miro and Kossack, Daniel and Campos, Simeon and Saouma, Lama and Chaudhry, Uzma and Soder, Lisa and Stein, Merlin and Caputo, Nicholas and Dunlop, Connor and Mökander, Jakob and Panai, Enrico and Lebrun, Tom and Martinet, Charles and Bucknall, Ben and Weiss, Rebecca and Holtman, Koen and Paskov, Patricia and Siddiqui, Saad and Barez, Fazl and Zuhdi, Ranj and Slattery, Peter and Ostmann, Florian},
  title = {Safety Frameworks and Standards: A comparative analysis to advance risk management of frontier AI},
  url = {https://aigi.ox.ac.uk/wp-content/uploads/2025/10/Post-convening-memo_-Safety-Frameworks-and-Standards_-A-comparative-analysis-to-advance-risk-management-of-frontier-AI_09.10.2025.pdf},
  urldate = {2025-10-10},
  year = {2025}
}

@misc{simpson_2025_voluntary,
  author = {Simpson, Morgan and Ortega, Alejandro and Trager, Robert},
  title = {Voluntary Industry Initiatives in Frontier AI Governance: Lessons…},
  url = {https://www.oxfordmartin.ox.ac.uk/publications/voluntary-industry-initiatives-in-frontier-ai-governance-lessons-from-aviation-and-nuclear-power},
  urldate = {2025-10-10},
  year = {2025},
  organization = {Oxford Martin School}
}

@article{lintern_2022_emergence,
  author = {Lintern, Gavan and Kugler, Peter N.},
  month = {10},
  pages = {1-16},
  title = {Emergence and non-emergence for system safety},
  doi = {10.1080/1463922x.2022.2134941},
  urldate = {2022-12-09},
  volume = {24},
  year = {2022},
  journal = {Theoretical Issues in Ergonomics Science}
}

@misc{wei_2022_emergent,
  author = {Wei, Jason and Tay, Yi and Bommasani, Rishi and Raffel, Colin and Zoph, Barret and Borgeaud, Sebastian and Yogatama, Dani and Bosma, Maarten and Zhou, Denny and Metzler, Donald and Chi, Ed and Hashimoto, Tatsunori and Vinyals, Oriol and Liang, Percy and Dean, Jeff and Fedus, William},
  month = {08},
  title = {Emergent Abilities of Large Language Models},
  url = {https://openreview.net/pdf?id=yzkSU5zdwD},
  year = {2022}
}

@article{harding_2025_what,
  author = {Harding, Jacqueline and Kirk-Giannini, Cameron Domenico},
  month = {06},
  publisher = {Springer Science+Business Media},
  title = {What is AI safety? What do we want it to be?},
  doi = {10.1007/s11098-025-02367-z},
  volume = {182},
  year = {2025},
  journal = {Philosophical Studies}
}

@misc{chen_2025_enhancing,
  author = {Chen, Yanda and Tucker, Mycal  and Panickssery, Nina  and Wang, Tony and Mosconi, Francesco and Gopal, Anjali and Denison, Carson and Petrini, Linda and Leike, Jan  and Perez, Ethan  and Sharma, Mrinank},
  title = {Enhancing Model Safety through Pretraining Data Filtering},
  url = {https://alignment.anthropic.com/2025/pretraining-data-filtering/},
  urldate = {2025-10-10},
  year = {2025},
  organization = {Anthropic.com}
}

@misc{casper_2023_open,
  author = {Casper, Stephen and Davies, Xander and Shi, Claudia and Gilbert, Thomas Krendl and Scheurer, Jérémy and Rando, Javier and Freedman, Rachel and Tomek Korbak and Lindner, David and Freire, Pedro and Wang, Tony Tong and Marks, Samuel and Charbel-Raphael Segerie and Carroll, Micah and Peng, Andi and Christoffersen, Phillip J.K and Damani, Mehul and Slocum, Stewart and Anwar, Usman and Anand Siththaranjan and Nadeau, Max and Michaud, Eric J and Pfau, Jacob and Dmitrii Krasheninnikov and Chen, Xin and Lauro Langosco and Hase, Peter and Erdem Biyik and Dragan, Anca and Krueger, David and Sadigh, Dorsa and Hadfield-Menell, Dylan},
  title = {Open Problems and Fundamental Limitations of Reinforcement Learning from Human Feedback},
  url = {https://openreview.net/forum?id=bx24KpJ4Eb},
  year = {2023},
  organization = {Transactions on Machine Learning Research}
}

@misc{guan_2024_deliberative,
  author = {Guan, Melody Y and Joglekar, Manas and Wallace, Eric and Jain, Saachi and Barak, Boaz and Helyar, Alec and Dias, Rachel and Vallone, Andrea and Ren, Hongyu and Wei, Jason and Chung, Hyung Won and Toyer, Sam and Heidecke, Johannes and Beutel, Alex and Glaese, Amelia},
  title = {Deliberative Alignment: Reasoning Enables Safer Language Models},
  url = {https://arxiv.org/abs/2412.16339?},
  urldate = {2025-10-10},
  year = {2024},
  organization = {arXiv.org}
}

@misc{sharma_2023_towards,
  author = {Sharma, Mrinank and Tong, Meg and Korbak, Tomasz and Duvenaud, David and Askell, Amanda and Bowman, Samuel R. and Cheng, Newton and Durmus, Esin and Hatfield-Dodds, Zac and Johnston, Scott R. and Kravec, Shauna and Maxwell, Timothy and McCandlish, Sam and Ndousse, Kamal and Rausch, Oliver and Schiefer, Nicholas and Yan, Da and Zhang, Miranda and Perez, Ethan},
  month = {10},
  title = {Towards Understanding Sycophancy in Language Models},
  doi = {10.48550/arXiv.2310.13548},
  url = {https://arxiv.org/abs/2310.13548},
  year = {2023},
  organization = {arXiv.org}
}

@misc{haddoncavekc_the,
  author = {Haddon-Cave KC, Charles},
  title = {The Nimrod Review: An independent review into the broader issues surrounding the loss of the RAF Nimrod MR2 Aircraft XV230 in Afghanistan in 2006},
  url = {https://assets.publishing.service.gov.uk/media/5a7c652640f0b62aff6c1609/1025.pdf},
  year = {2009}
}

@misc{fronsdal_2025_petri,
  author = {Fronsdal, Kai and Gupta, Isha and Sheshadri, Abhay and Michala, Jonathan and McAleer, Stephen and Wang, Rowan and Price, Sara and Bowman, Samuel},
  title = {Petri: An open-source auditing tool to accelerate AI safety research},
  url = {https://alignment.anthropic.com/2025/petri/},
  urldate = {2025-10-10},
  year = {2025},
  organization = {Anthropic.com}
}

@misc{kalai_2025_why,
  author = {Kalai, Adam and Nachum, Ofir and Vempala, Santosh and Zhang, Edwin},
  title = {Why Language Models Hallucinate},
  url = {https://cdn.openai.com/pdf/d04913be-3f6f-4d2b-b283-ff432ef4aaa5/why-language-models-hallucinate.pdf},
  year = {2025}
}

@misc{openai_2025_gpt5,
  author = {OpenAI},
  title = {GPT-5 System Card},
  url = {https://cdn.openai.com/gpt-5-system-card.pdf},
  year = {2025}
}

@misc{googledeepmind_2025_gemini,
  author = {Google DeepMind},
  title = {Gemini 2.5 Deep Think Model Card},
  url = {https://storage.googleapis.com/deepmind-media/Model-Cards/Gemini-2-5-Deep-Think-Model-Card.pdf},
  year = {2025}
}

@misc{anthropic_2025_system,
  author = {Anthropic},
  title = {System Card: Claude Sonnet 4.5},
  url = {https://assets.anthropic.com/m/12f214efcc2f457a/original/Claude-Sonnet-4-5-System-Card.pdf},
  year = {2025}
}

@misc{ministryofdefence_2025_hazard,
  author = {Ministry of Defence},
  publisher = {Ministry of Defence},
  title = {Hazard Log: Acquisition Safety \& Environmental Management System},
  url = {https://www.asems.mod.uk/toolkit/hazard-log},
  urldate = {2025-10-10},
  year = {2025},
  organization = {Asems.mod.uk}
}

@inproceedings{alexander_2008_engineering,
  author = {Alexander, Robert and Alexander-Brown, Ruth and Kelly, Timothy},
  pages = {33-62},
  publisher = {Luniver Press},
  title = {Engineering Safety-Critical Complex Systems},
  year = {2008},
  booktitle = {CoSMoS 2008: Proceedings of the 2008 Workshop on Complex Systems Modelling and Simulation }
}

@misc{departmentforscienceinnovationandtechnology_2023_capabilities,
  author = {Department for Science, Innovation and Technology},
  title = {Capabilities and risks from frontier AI},
  url = {https://assets.publishing.service.gov.uk/media/65395abae6c968000daa9b25/frontier-ai-capabilities-risks-report.pdf},
  year = {2023}
}

@misc{buhl_2024_safety,
  author = {Buhl, Marie Davidsen and Sett, Gaurav and Koessler, Leonie and Schuett, Jonas and Anderljung, Markus},
  month = {10},
  publisher = {Cornell University},
  title = {Safety cases for frontier AI},
  doi = {10.48550/arxiv.2410.21572},
  urldate = {2025-01-21},
  year = {2024},
  journal = {arXiv}
}

@misc{clymer_2024_safety,
  author = {Clymer, Joshua and Gabrieli, Nick and Krueger, David and Larsen, Thomas},
  title = {Safety Cases: How to Justify the Safety of Advanced AI Systems},
  url = {https://arxiv.org/abs/2403.10462},
  year = {2024},
  organization = {arXiv.org}
}

@misc{hilton_2025_safety,
  author = {Hilton, Benjamin and Buhl, Marie Davidsen and Korbak, Tomek and Irving, Geoffrey},
  title = {Safety Cases: A Scalable Approach to Frontier AI Safety},
  url = {https://arxiv.org/abs/2503.04744},
  urldate = {2025-07-09},
  year = {2025},
  organization = {arXiv.org}
}

@misc{anthropic_2024_three,
  author = {Anthropic},
  title = {Three Sketches of ASL-4 Safety Case Components},
  url = {https://alignment.anthropic.com/2024/safety-cases/},
  year = {2024},
  organization = {Anthropic.com}
}

@misc{korbak_2025_a,
  author = {Korbak, Tomek and Clymer, Joshua and Hilton, Benjamin and Shlegeris, Buck and Irving, Geoffrey},
  title = {A sketch of an AI control safety case},
  url = {https://arxiv.org/abs/2501.17315},
  urldate = {2025-07-09},
  year = {2025},
  organization = {arXiv.org}
}

@misc{buhl_2025_an,
  author = {Buhl, Marie Davidsen and Pfau, Jacob and Hilton, Benjamin and Irving, Geoffrey},
  title = {An alignment safety case sketch based on debate},
  url = {https://arxiv.org/abs/2505.03989},
  year = {2025},
  organization = {arXiv.org}
}

@misc{goemans_2024_safety,
  author = {Goemans, Arthur and Buhl, Marie Davidsen and Schuett, Jonas and Korbak, Tomek and Wang, Jessica and Hilton, Benjamin and Irving, Geoffrey},
  title = {Safety case template for frontier AI: A cyber inability argument},
  url = {https://arxiv.org/abs/2411.08088},
  year = {2024},
  organization = {arXiv.org}
}

@misc{balesni_2024_towards,
  author = {Balesni, Mikita and Hobbhahn, Marius and Lindner, David and Meinke, Alexander and Korbak, Tomek and Clymer, Joshua and Shlegeris, Buck and Scheurer, Jérémy and Stix, Charlotte and Shah, Rusheb and Goldowsky-Dill, Nicholas and Braun, Dan and Chughtai, Bilal and Evans, Owain and Kokotajlo, Daniel and Bushnaq, Lucius},
  title = {Towards evaluations-based safety cases for AI scheming},
  url = {https://arxiv.org/abs/2411.03336},
  urldate = {2025-02-13},
  year = {2024},
  organization = {arXiv.org}
}

@misc{ortega_2025_ai,
  author = {Ortega, Alejandro},
  title = {AI threats to national security can be countered through an incident regime},
  url = {https://arxiv.org/abs/2503.19887},
  year = {2025},
  organization = {arXiv.org}
}

@misc{clymer_2025_an,
  author = {Clymer, Joshua and Weinbaum, Jonah and Kirk, Robert and Mai, Kimberly and Zhang, Selena and Davies, Xander},
  title = {An Example Safety Case for Safeguards Against Misuse},
  url = {https://arxiv.org/abs/2505.18003},
  urldate = {2025-10-10},
  year = {2025},
  organization = {arXiv.org}
}

@misc{korbak_2025_how,
  author = {Korbak, Tomek and Balesni, Mikita and Shlegeris, Buck and Irving, Geoffrey},
  title = {How to evaluate control measures for LLM agents? A trajectory from today to superintelligence},
  url = {https://arxiv.org/abs/2504.05259},
  year = {2025},
  organization = {arXiv.org}
}

@misc{barrett_2025_assessing,
  author = {Barrett, Stephen and Fox, Philip and Krook, Joshua and Mondal, Tuneer and Mylius, Simon and Tlaie, Alejandro},
  title = {Assessing confidence in frontier AI safety cases},
  url = {https://arxiv.org/abs/2502.05791},
  urldate = {2025-10-10},
  year = {2025},
  organization = {arXiv.org}
}

@article{alexander_2017_from,
  author = {Alexander, Robert David and Hawkins, Richard David and Kelly, Timothy Patrick},
  month = {02},
  title = {From Safety Cases to Security Cases},
  urldate = {2025-10-10},
  year = {2017},
  journal = {Safety Critical Systems Symposium 2017}
}

@misc{hao_2024_training,
  author = {Hao, Shibo and Sukhbaatar, Sainbayar and Su, DiJia and Li, Xian and Hu, Zhiting and Weston, Jason and Tian, Yuandong},
  title = {Training Large Language Models to Reason in a Continuous Latent Space},
  url = {https://arxiv.org/abs/2412.06769},
  urldate = {2025-01-13},
  year = {2024},
  journal = {arXiv.org}
}

@misc{korbak_2025_chain,
  author = {Korbak, Tomek and Balesni, Mikita and Barnes, Elizabeth and Bengio, Yoshua and Benton, Joe and Bloom, Joseph and Chen, Mark and Cooney, Alan and Dafoe, Allan and Dragan, Anca and Emmons, Scott and Evans, Owain and Farhi, David and Greenblatt, Ryan and Hendrycks, Dan and Hobbhahn, Marius and Hubinger, Evan and Irving, Geoffrey and Jenner, Erik and Kokotajlo, Daniel and Krakovna, Victoria and Legg, Shane and Lindner, David and Luan, David and Mądry, Aleksander and Michael, Julian and Nanda, Neel and Orr, Dave and Pachocki, Jakub and Perez, Ethan and Phuong, Mary and Roger, Fabien and Saxe, Joshua and Shlegeris, Buck and Soto, Martín and Steinberger, Eric and Wang, Jasmine and Zaremba, Wojciech and Baker, Bowen and Shah, Rohin and Mikulik, Vlad},
  title = {Chain of Thought Monitorability: A New and Fragile Opportunity for AI Safety},
  url = {https://arxiv.org/abs/2507.11473v1},
  urldate = {2025-10-10},
  year = {2025},
  organization = {arXiv.org}
}

@article{kelly_2008_are,
  author = {Kelly, Timothy},
  pages = {31-33},
  title = {Are safety cases working?},
  volume = {17},
  year = {2008},
  journal = {Safety Critical Systems Club Newsletter}
}

@misc{internationalstandardsorganisation_2014_isoiec,
  author = {International Standards Organisation},
  publisher = {International Standards Organisation},
  title = {ISO-IEC 51-2014},
  url = {https://www.iso.org/standard/53940.html},
  year = {2014}
}

@inproceedings{FearnleyForthcoming-FEACCI-2,
	author = {Laura Fearnley and Ibrahim Habli},
	booktitle = {Proceedings of the Eight Aaai/Acm Conference on Ai, Ethics, and Society (Aies-25)},
	title = {Concept Creep in Safe Artificial Intelligence},
	year = {2025}
}

@misc{davies_2023_regulating,
  author = {Davies, Matt and Birtwistle, Michael},
  month = {07},
  title = {Regulating AI in the UK},
  url = {https://www.adalovelaceinstitute.org/report/regulating-ai-in-the-uk/},
  year = {2023},
  organization = {www.adalovelaceinstitute.org}
}

@article{redmill_1999_understanding,
  author = {Redmill, Felix},
  month = {09},
  pages = {197-200},
  title = {Understanding Safety Integrity Levels},
  doi = {10.1177/002029409903200702},
  url = {https://pdfs.semanticscholar.org/390c/b437b7995d5d44053efe56746bcf1d846e52.pdf},
  urldate = {2019-03-09},
  volume = {32},
  year = {1999},
  journal = {Measurement and Control}
}

@inproceedings{mcdermid_2001_scs,
  author = {McDermid, John},
  title = {Software safety: where's the evidence?
},
  pages = {1-6},
  booktitle = {SCS '01 Proceedings of the Sixth Australian workshop on Safety critical systems and software},
  volume = {1},
  year = {2001},
  organization = {SCS}
}

@article{littlewood_1993_validation,
  author = {Littlewood, Bev and Strigini, Lorenzo},
  month = {11},
  pages = {69-80},
  title = {Validation of ultrahigh dependability for software-based systems},
  doi = {10.1145/163359.163373},
  url = {https://dl.acm.org/citation.cfm?doid=163359.163373},
  urldate = {2019-08-13},
  volume = {36},
  year = {1993},
  journal = {Communications of the ACM}
}

@misc{healthandsafetyexecutive_2001_reducing,
  author = {Health and Safety Executive},
  title = {Reducing risks, HSE's decision-making process protecting people},
  url = {https://assets.publishing.service.gov.uk/media/6693ad9e49b9c0597fdafc36/IQ8.10.J_Document_9_Health_and_Safety_Executive__Reducing_risks__protecting_people__HSE_s_decision-making_process__2001.pdf},
  year = {2001}
}

@misc{researchexcellenceframework_2014_com04,
  author = {Research Excellence Framework},
  title = {COM04 The Goal Structuring Notation (GSN},
  url = {https://impact.ref.ac.uk/casestudies/CaseStudy.aspx?Id=43445},
  urldate = {2025-10-10},
  year = {2014},
  organization = {Ref.ac.uk}
}

@misc{goalstructuringnotationstandardworkinggroup_2011_gsn,
  author = {Goal Structuring Notation Standard Working Group},
  publisher = {Safety Critical Systems Club},
  title = {GSN COMMUNITY STANDARD VERSION 1},
  url = {https://scsc.uk},
  year = {2011}
}

@misc{irving_2024_safety,
  author = {Irving, Geoffrey},
  title = {Safety cases at AISI | AISI Work},
  url = {https://www.aisi.gov.uk/blog/safety-cases-at-aisi},
  urldate = {2025-10-10},
  year = {2024},
  organization = {AI Security Institute}
}

@misc{hendrycks_2023_an,
  author = {Hendrycks, Dan and Mazeika, Mantas and Woodside, Thomas},
  month = {06},
  title = {An Overview of Catastrophic AI Risks},
  doi = {10.48550/arXiv.2306.12001},
  url = {https://arxiv.org/abs/2306.12001},
  year = {2023},
  organization = {arXiv.org}
}

@misc{hubinger_2022_how,
  author = {Hubinger, Evan},
  month = {08},
  title = {How likely is deceptive alignment?},
  url = {https://www.lesswrong.com/posts/A9NxPTwbw6r6Awuwt/how-likely-is-deceptive-alignment},
  year = {2022},
  organization = {Lesswrong.com}
}

@misc{apolloresearch_2023_understanding,
  author = {Apollo Research},
  month = {09},
  title = {Understanding strategic deception and deceptive alignment – Apollo Research},
  url = {https://www.apolloresearch.ai/blog/understanding-strategic-deception-and-deceptive-alignment/},
  urldate = {2026-02-08},
  year = {2023},
  organization = {Apollo Research}
}

@inproceedings{carranza_2023_deceptive,
  author = {Carranza, Andres and Pai, Dhruv and Schaeffer, Rylan and Tandon, Arnuv and Koyejo, Sanmi},
  title = {Deceptive Alignment Monitoring},
  url = {https://openreview.net/pdf?id=obsO44GFhh},
  urldate = {2026-02-08},
  year = {2023},
  booktitle = {2023 ICML AdvML Workshop}
}

@inproceedings{zhou_2024_robust,
  author = {Zhou, Andy and Li, Bo and Wang, Haohan},
  month = {09},
  title = {Robust Prompt Optimization for Defending Language Models Against Jailbreaking Attacks},
  url = {https://neurips.cc/virtual/2024/poster/93953},
  urldate = {2026-02-08},
  year = {2024},
  booktitle = {Advances in Neural Information Processing Systems 38}
}

@misc{sharma_2025_constitutional,
  author = {Sharma, Mrinank and Tong, Meg and Mu, Jesse and Wei, Jerry and Kruthoff, Jorrit and Goodfriend, Scott and Ong, Euan and Peng, Alwin and Agarwal, Raj and Anil, Cem and Askell, Amanda and Bailey, Nathan and Benton, Joe and Bluemke, Emma and Bowman, Samuel R and Christiansen, Eric and Cunningham, Hoagy and Dau, Andy and Gopal, Anjali and Gilson, Rob and Graham, Logan and Howard, Logan and Kalra, Nimit and Lee, Taesung and Lin, Kevin and Lofgren, Peter and Mosconi, Francesco and O'Hara, Clare and Olsson, Catherine and Petrini, Linda and Rajani, Samir and Saxena, Nikhil and Silverstein, Alex and Singh, Tanya and Sumers, Theodore and Tang, Leonard and Troy, Kevin K and Weisser, Constantin and Zhong, Ruiqi and Zhou, Giulio and Leike, Jan and Kaplan, Jared and Perez, Ethan},
  title = {Constitutional Classifiers: Defending against Universal Jailbreaks across Thousands of Hours of Red Teaming},
  url = {https://arxiv.org/abs/2501.18837},
  year = {2025},
  organization = {arXiv.org}
}
\bibliographystyle{ACM-Reference-Format}


\appendix

\section{Technical Appendices and Supplementary Material}

\subsection{Appendix A: Full GSN walkthrough}

\begin{figure}[H]
    \centering
    \includegraphics[width=0.5\linewidth]{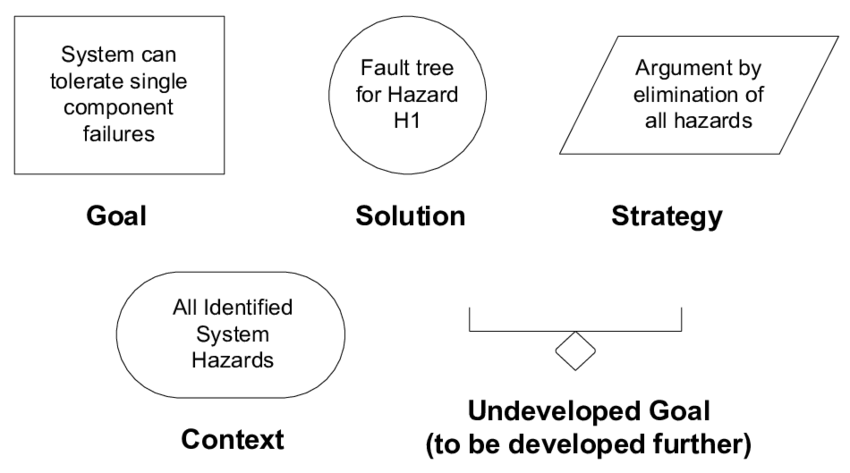}
    \caption{GSN Argument Blocks: \cite{alexander_2008_engineering}}
     \label{GSN Argument Blocks}
\end{figure}

\begin{figure}[H]
\includegraphics[width=0.5\linewidth]{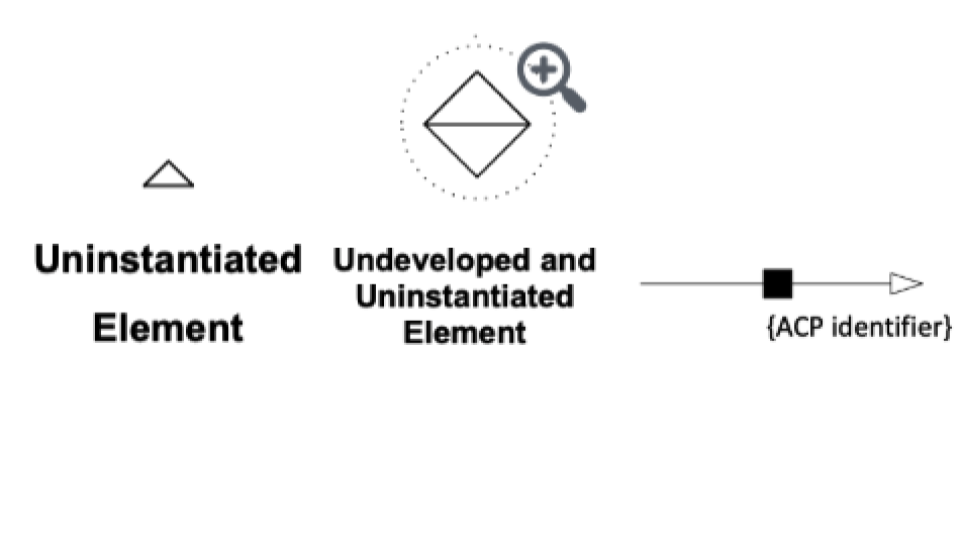}
    \centering
    \caption{ACPs are used to indicate that a claim is accompanied by a confidence assertion. Uninstantiated and undeveloped elements indicate goals or evidence which need to be completed with a more concrete instance. \cite{goalstructuringnotationstandardworkinggroup_2011_gsn}}
     \label{Elements in GSN}
\end{figure}

\begin{figure}[H]
    \centering
    \includegraphics[width=0.5\linewidth]{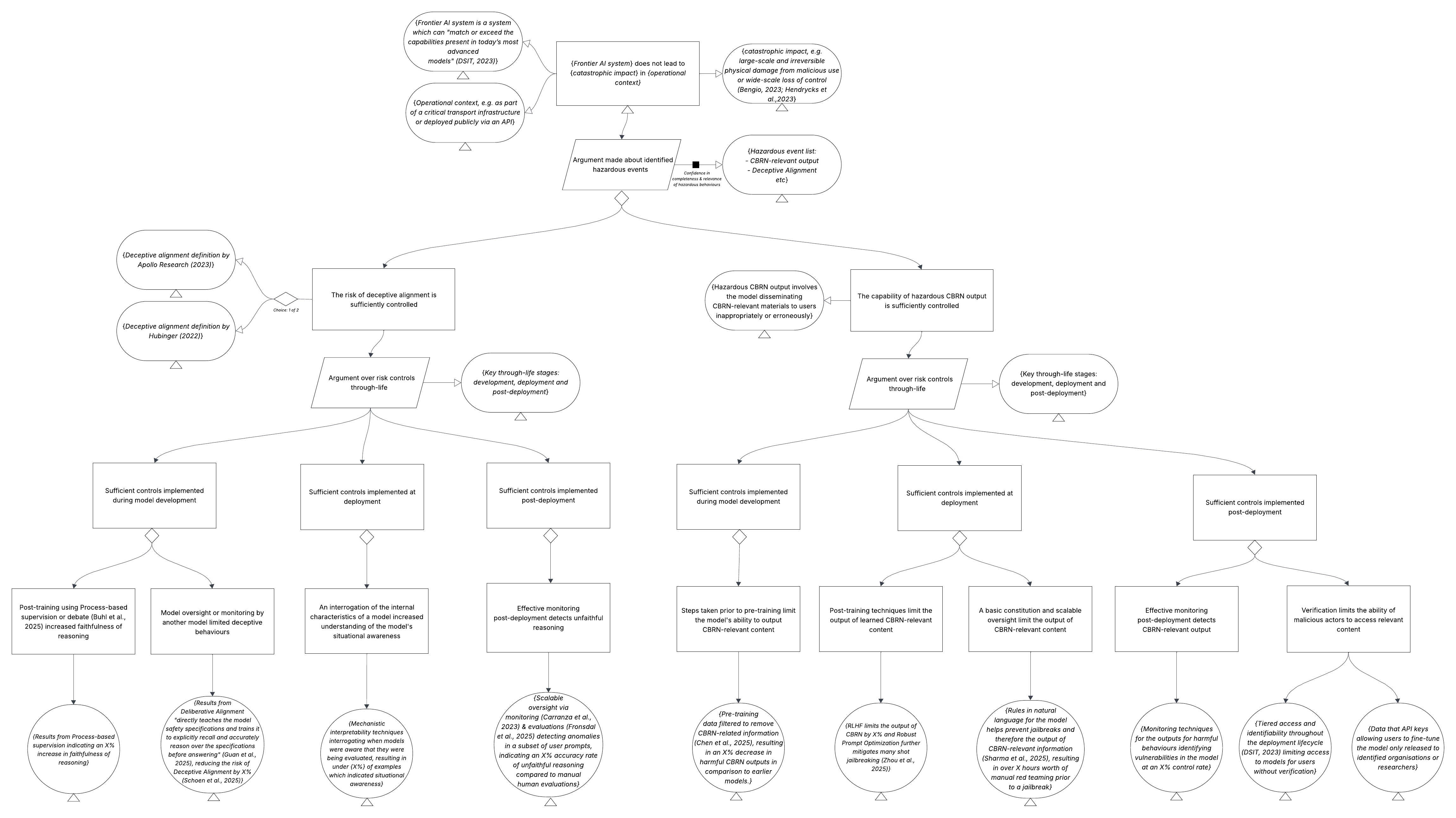}
    \caption{The full GSN can be traced from a top-level goal down to supporting evidence for relevant subgoals and strategies}
     \label{Full GSN}
\end{figure}

The safety argument begins with a top-level goal, accompanied by assertions which include defined terms:

\begin{figure}[H]
    \centering
    \includegraphics[width=0.5\linewidth]{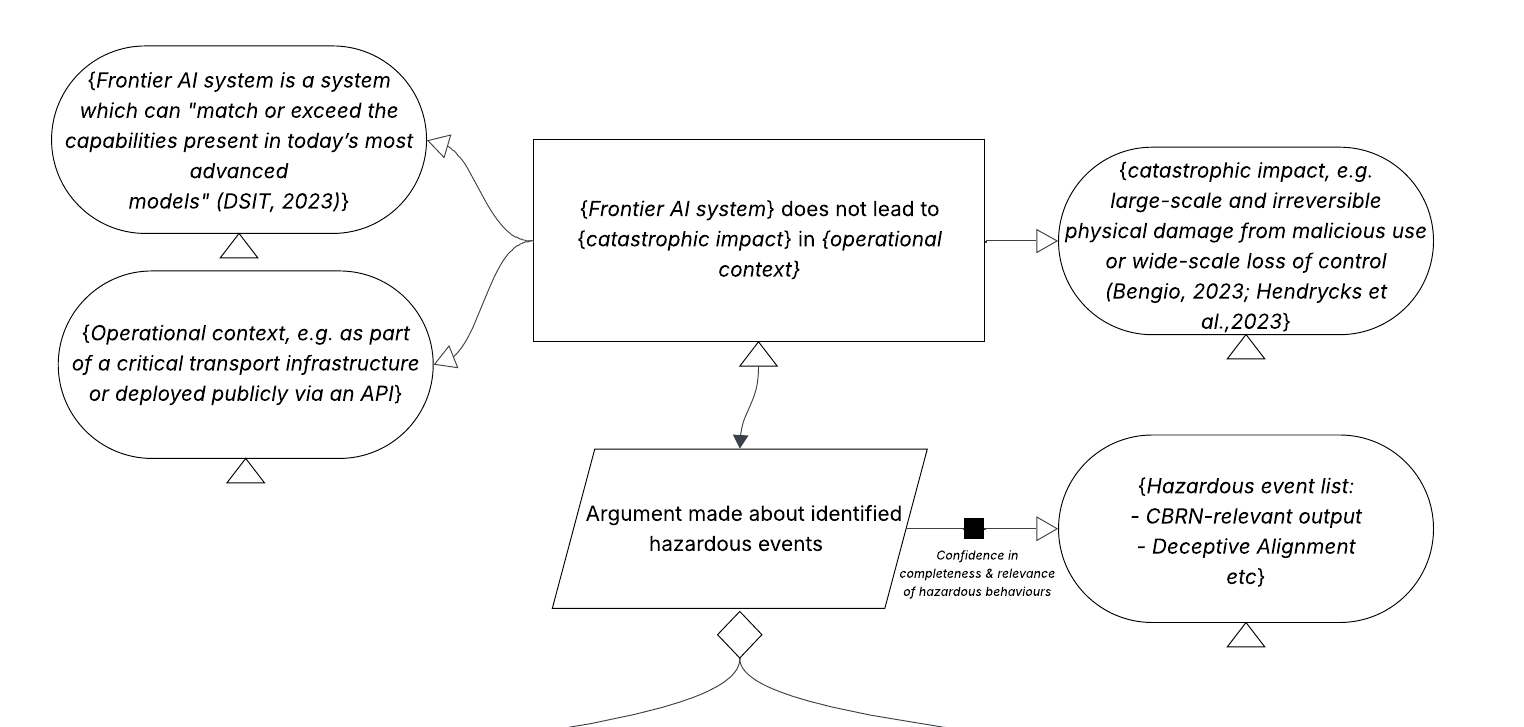}
    \caption{The top-level goal is supported by examination of identified hazardous events. Each hazardous event is accompanied by supporting goals and evidence, which is derived from an earlier risk assessment and risk reduction or evaluation methods. Evidence in GSN, i.e. solutions, are not claims but references to data or results.  We focus on two of many potentially hazardous events. This case study focuses firstly on Deceptive Alignment, a hazard which is defined with less certainty:}
    \label{Top-Level Goal}
\end{figure}
The argument then moves to sub-goals, which are supported by solutions in the form of specific evidence:
\begin{figure}[H]
    \centering
    \includegraphics[width=0.3\linewidth]{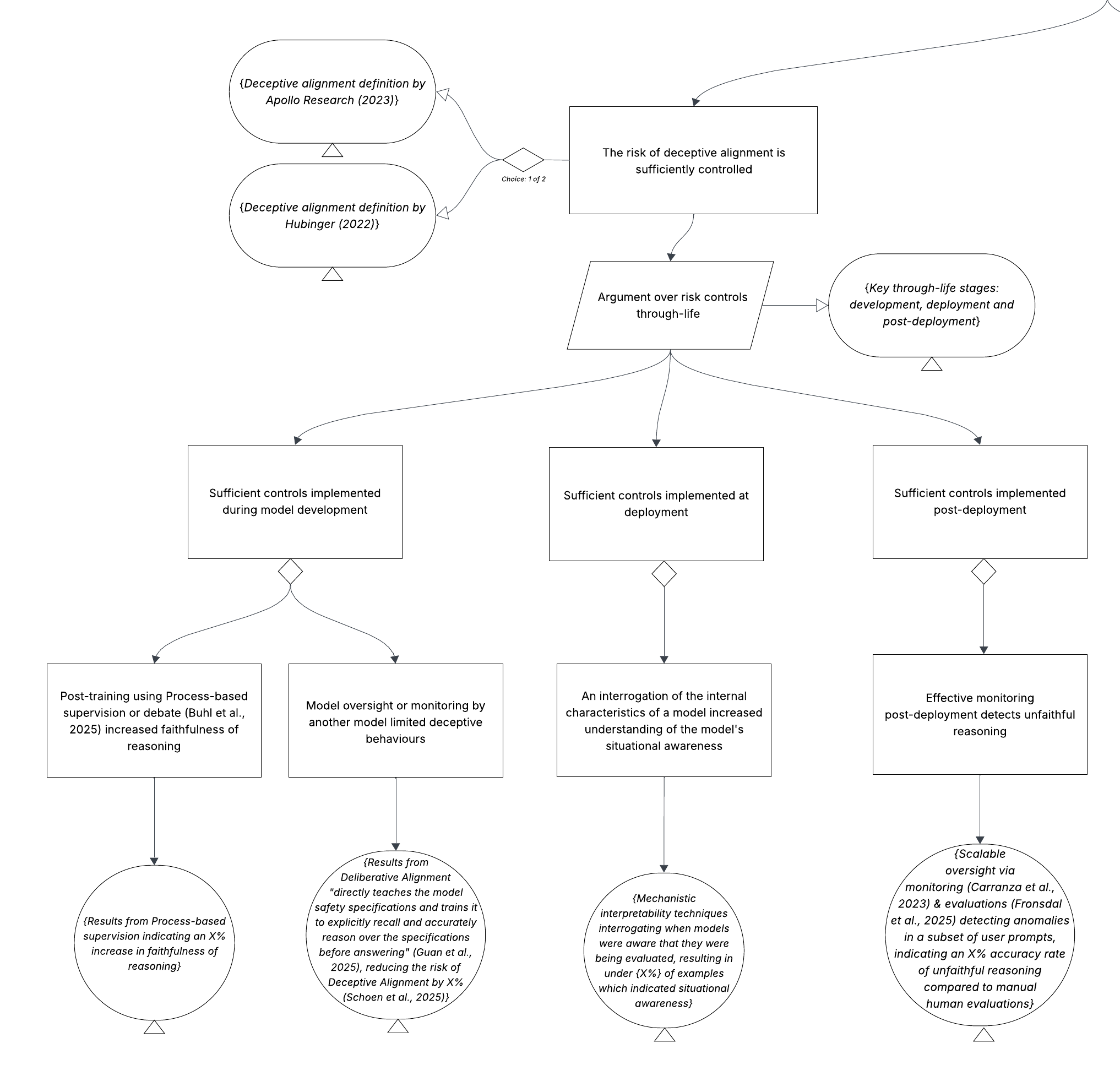}
    \caption{Control of the hazardous event is supported by arguments over through-life controls and mitigations, split here into development, deployment and post-deployment. Each goal is then supported by evidence}
    \label{Deceptive Alignment}
\end{figure}

\begin{figure}[H]
    \centering
    \includegraphics[width=0.3\linewidth]{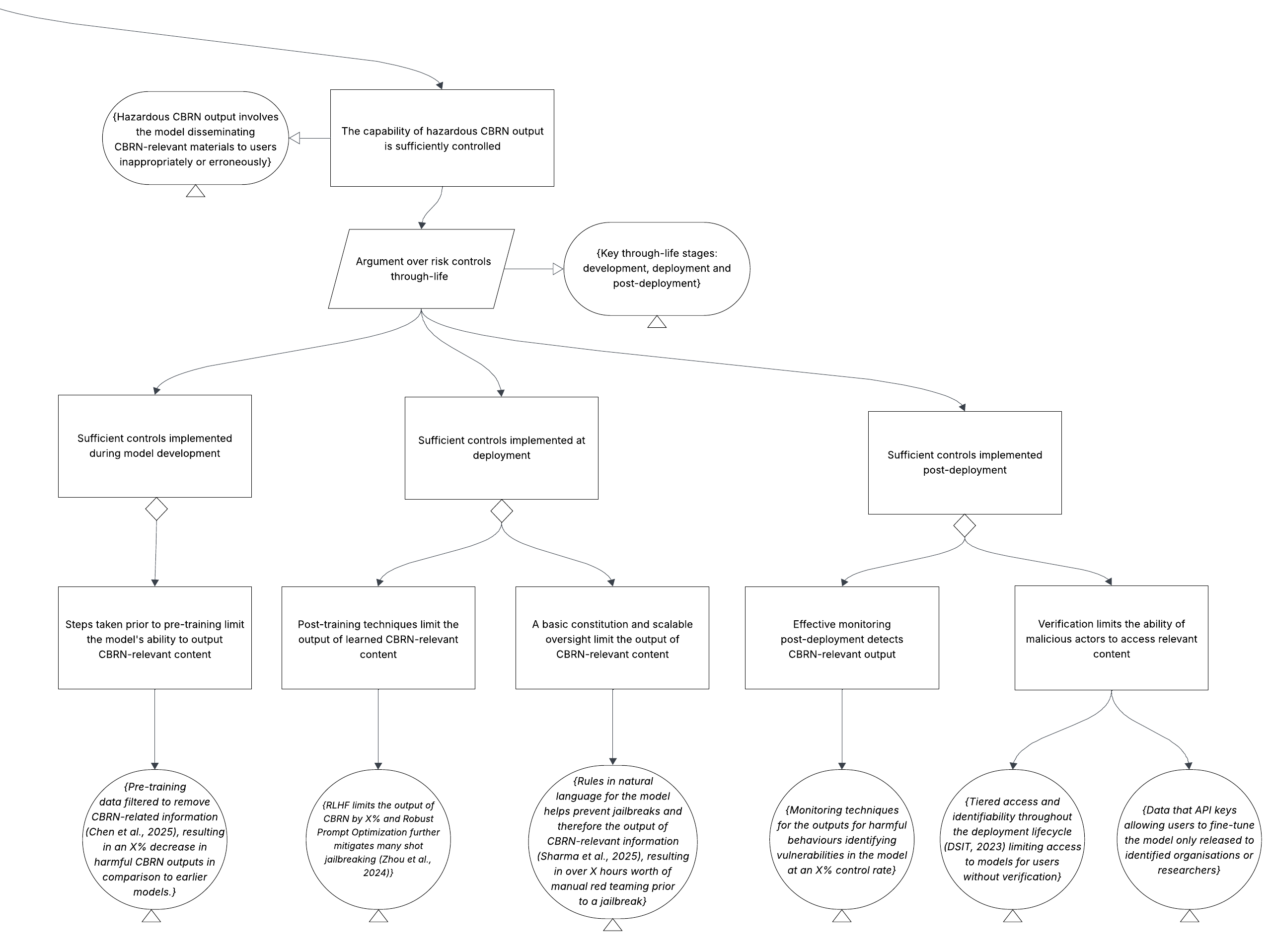}
    \caption{Control of the hazardous event is supported by arguments over through-life controls and mitigations, split here into development, deployment and post-deployment. Each goal is then supported by evidence}
    \label{Capability Hazard}
\end{figure}

\subsection{Appendix B: Alignment Safety Case Research}

Illustrative list of relevant safety case documents or research produced by the U.K. AI Security Institute, Apollo Research, Redwood Research and Frontier AI companies, or cited by the International AI Safety Report. No cited work on alignment safety cases has undergone formal peer review, likely due to the short timeframes within which the frontier AI community operates. The relevant source relates to the correspondence email or publishing website of the research:

\textit{Title: }{Safety Cases for Frontier AI}{, 2024} \\
  \textit{Authors:} MD Buhl, G Sett, L Koessler, J Schuett, M Anderljung\\
  \textit{Source:} Centre for Governance of AI \cite{buhl_2024_safety}

\textit{Title: }{Safety Cases: How to Justify the Safety of Advanced AI Systems}{, 2024} \\
  \textit{Authors:} J Clymer, N Gabrieli, D Krueger, T Larsen\\
  \textit{Source:} Independent \cite{clymer_2024_safety}

\textit{Title: }{Safety Cases: A Scalable Approach to Frontier AI Safety}{, 2025} \\
  \textit{Authors:} B Hilton, MD Buhl, T Korbak, G Irving\\
  \textit{Source:} U.K. AI Security Institute \cite{hilton_2025_safety}

\textit{Title: }{Three Sketches of ASL-4 Safety Case Components}{, 2025} \\
  \textit{Authors:} R Grosse\\
  \textit{Source:} Anthropic \cite{anthropic_2024_three}

\textit{Title: }{A sketch of an AI control safety case}{, 2025} \\
  \textit{Authors:} T Korbak, J Clymer, B Hilton, B Shlegeris, G Irving\\
  \textit{Source:} Redwood Research, U.K. AI Security Institute \cite{korbak_2025_a}

Title: {An alignment safety case sketch based on debate}{, 2025} \\
  \textit{Authors:} MD Buhl, J Pfau, B Hilton, G Irving\\
  \textit{Source:} AI Security Institute \cite{buhl_2025_an}

Title: {Safety Case Template for Frontier AI: A Cyber Inability Argument}{, 2025} \\
  \textit{Authors:} A Goemans, MD Buhl, J Schuett, T Korbak, J Wang, B Hilton, G Irving\\
  \textit{Source:} Centre for Governance of AI \cite{goemans_2024_safety}

\textit{Title: }{Towards evaluations-based safety cases for AI scheming}{, 2025} \\
  \textit{Authors: }M Balesni, M Hobbhahn, D Lindner, A Meinke, T Korbak, J Clymer, B Shlegeris, J Scheurer, C Stix, R Shah, N Goldwosky-Dill, D Braun, B Chughtai, O Evans, D Kokotajlo, L Bushnaq\\
  \textit{Source:} Apollo Research \cite{balesni_2024_towards}

\textit{Title: }{AI threats to national security can be countered through an incident regime}{, 2025} \\
  \textit{Authors:} A Ortega\\
  \textit{Source:} Apollo Research \cite{ortega_2025_ai}

\textit{Title: }{An Example Safety Case for Safeguards Against Misuse}{, 2025} \\
  \textit{Authors:} J Clymer, J Weinbaum, R Kirk, K Mai, S Zhang, X Davies\\
  \textit{Source:} Independent, U.K. AI Security Institute \cite{clymer_2025_an}

\textit{Title: }{How to evaluate control measures for LLM agents? A trajectory from today to superintelligence}{, 2025} \\
  \textit{Authors:} T Korbak, M Balesni, B Shlegeris, G Irving\\
  \textit{Source:} U.K. AI Security Institute \cite{korbak_2025_how}

Title: {Responsible Scaling Policy}{, 2025} \\
  \textit{Source:} Anthropic \cite{anthropic_2024_responsible}

\textit{Title: }{Frontier Safety Framework 2.0}{, 2025} \\
  \textit{Source:} Google DeepMind \cite{googledeepmind_2025_frontier}

The list is intended to be illustrative rather than comprehensive, particularly given the fast-paced nature of the field.

\subsection{Appendix C: Risk Assessment Workflow}

\begin{figure}[H]
    \centering
    \includegraphics[width=0.8\linewidth]{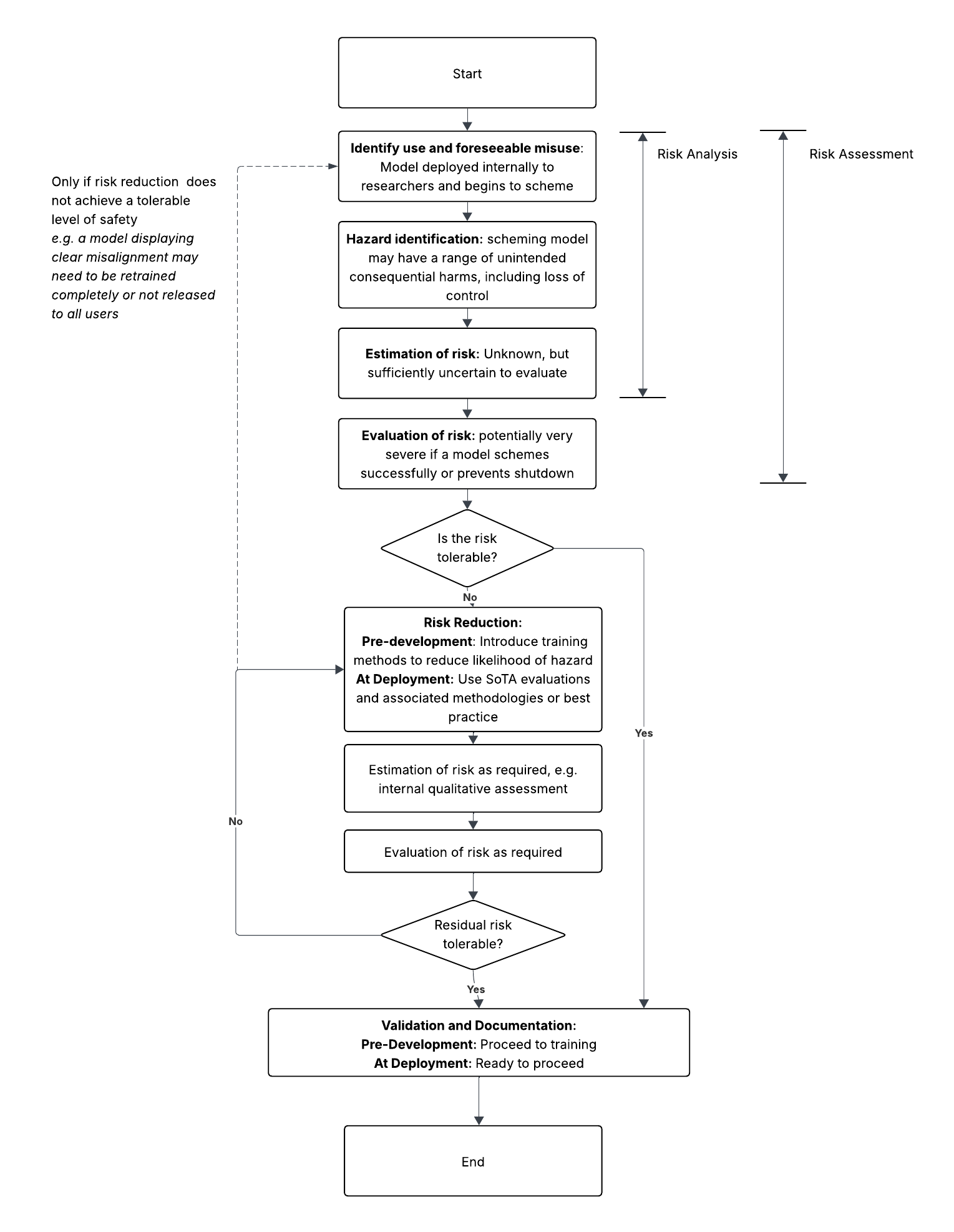}
    \caption{Risk Reduction Workflow \cite{internationalstandardsorganisation_2014_isoiec}}
    \label{Fig 2: Risk Reduction Workflow}
\end{figure}


\end{document}